\documentclass[aps,pra,superscriptaddress,twocolumn]{revtex4}
\usepackage[utf8x]{inputenc}
\usepackage{float}
\usepackage{bm}
\usepackage{graphicx,amsmath,latexsym,amssymb}
\usepackage{color}
\usepackage{dcolumn}

\def\bra#1{\mathinner{\langle{#1}|}}
\def\ket#1{\mathinner{|{#1}\rangle}}

\def\Bra#1{\left<1>}
{\catcode`\|=\active\gdef\Braket#1{\left<\mathcode`\|"8000\let|\bravert {#1}\right>}}
\def\bravert{\egroup\,\vrule\,\bgroup}

\begin{document}
\title{Resonant scattering of a single atom with gain: a wavefunction-diagrammatic approach}
\author{M. Donaire}
\email{manuel.donaire@uva.es}
\affiliation{Departamento de F\'isica Te\'orica, At\'omica y \'Optica and IMUVA,  Universidad de Valladolid, Paseo Bel\'en 7, 47011 Valladolid, Spain}

\begin{abstract}
We characterize the optical response of a three-level atom subjected to an incoherent pump and continuously illuminated with a weak, quasi-resonant probe field. 
To this end, we apply a wavefunction approach based on QED Hamiltonian perturbation theory which allows for a reduction of the atomic dynamics to that of an 
effective two-level atom, and for an  implementation of 
the incoherent effects that respects unitarity. 
Using a diagrammatic representation, we identify and classify all the radiative processes. 
This allows us to compute the scattered power, the spontaneous emission, and the stimulated emission, as well as the total cross sections of extinction, 
absorption and scattering. We find that, beside a general enhancement of the linewidth and an attenuation of the spectral amplitudes, the pump reduces the nonradiative losses and provides 
gains in the form of stimulated emission and incoherent radiation. For sufficiently strong pump, gains and losses compensate, resulting in the vanishing of extinction. In particular, for negligible nonradiative losses, extinction vanishes for a pumping rate of $(1+\sqrt{5})/2$ times that of the natural decay.

\end{abstract}

\maketitle

\section{Introduction and Motivation}

The scattering of light by a free atomic dipole in its ground state is a subject extensively studied within the standard QED theory, in any regime of physical interest 
\cite{Sakurai,Peskin_book,Milonni_book}. On the other hand, in order to describe  the scattering properties of a multi-level atom semiclassically, it is also possible to characterize its optical response  with an effective polarizability \cite{Milonni-Berman,Loudon-Berman,vanTiggelen,Buhmann-Scheel,Coevorden}, 
$\alpha(\omega)$, within the framework of the 
linear response theory and the $T$-matrix formalism. Operating this way, radiative dissipation appears parametrized by a term proportional to the imaginary part of the electromagnetic (EM) field propagator, Im$\{\mathbb{G}\}$, whereas its real part 
Re$\{\mathbb{G}\}$ is related to a shift in the atomic transition frequency 
analogous to that of the Lamb shift \cite{vanTiggelen,Buhmann-Scheel,Coevorden,Donaire,WyleySipe}. In addition,  nonradiative losses are accounted for  in an effective manner by adding an imaginary damping parameter on top of that proportional to  Im$\{\mathbb{G}\}$, in such a way that the resultant 
polarizability is compatible with the optical theorem. When illuminated by an external field of frequency $\omega$, $\mathbf{E}_{0}(\omega)$, the nature of the radiation scattered by the atom is  
interpreted on the basis of its classical response  to the external field. Thus, it is customary to referred as 
coherent scattered power (\emph{coh}) to that in phase with the expectation value of the atomic dipole moment, 
$\langle\mathbf{d}(\omega)\rangle=\alpha(\omega)\cdot\mathbf{E}_{0}(\omega)$, 
$\mathcal{W}_{coh}\sim$Im$\{\omega\langle\mathbf{d}(\omega)\rangle\cdot\mathbf{E}^{*}(\omega)\}$, with $\mathbf{E}(\omega)$ being the field radiated by the dipole 
itself \cite{Novotny}. Further, when the atom is coherently driven by external fields, a non-linear polarizability can be computed from the density-matrix formalism \cite{OBrien,Hang,Scully}. Finally, in regards to ensembles of atomic dipoles,  their internal energy as well as their collective optical response are derived out of the single-atom polarizabilities applying multiple scattering  techniques within the framework of the linear response theory \cite{Bullough,Heyne-Bullough,Donaire,Agarwal,Buhmann-Wess,vanTiggelen-Lagendij_Phys_Rept,vanTiggelen,deVries-Lagendij,Salam}.

As for an optical medium with incoherent gain, it is tempting to implement the gains by adding an imaginary damping term to its semiclassical dielectric response, of opposite 
sign with respect to that of the losses. While this  procedure is followed in some semiclassical systems --cf. Refs.\cite{Khandekar,ManjavacasPT}, it cannot  
be applied to the case of an atomic medium --eg., the active medium of a laser subjected to an incoherent pump. The reason being that, while semiclassical media support boson-like excitations, the atomic excitations behave as fermions in the sense that the population rates of the atomic states lie within the interval $[0,1]$. In turn this implies that the period of coherent evolution increases with the gain for a boson-like system, while it decreases 
for an atomic system \cite{TejedorII}. This is confirmed by the computation of the effective Bloch equations of a three-level atom subjected to an incoherent pump, which can be reduced to those of an effective two-level system integrating out the dynamics of the unstable upper state \cite{German,Lagendijk_3_level}.  Further, if that atom is illuminated with an additional 
probe field $\mathbf{E}_{0}(\omega)$, an effective polarizability $\alpha$ can be derived from the quantum computation of the induced dipole moment, $\langle\mathbf{d}(\omega)\rangle=\alpha(\omega)\cdot\mathbf{E}_{0}(\omega)$ \cite{Lagendijk_3_level}. 
Next, the computation of the  scattered power lies in the application of the aforementioned classical expression for $\mathcal{W}_{coh}$ \cite{Novotny},  
or in the application of the   quantum regression theorem upon the dipole moment quadratic fluctuations \cite{Scully,Carmichel,Harry_Paul,Hertel-Schulz}. 
In either case only the dynamics of the atomic states is treated quantum-mechanically, while the scattered power is a derivative product in which the actual 
nature of the radiation is generally unclear \cite{footnote}. As a result, odd results like the vanishing of the scattering cross-section are obtained 
\cite{Lagendijk_3_level}.

The aim of this article is to develop a wavefunction diagrammatic approach to study the optical response of a three-level atom to a quasi-resonant probe field 
while subjected to incoherent pumping. To this end we apply time-dependent Hamiltonian  perturbation theory, treating the atomic and photonic degrees of freedom 
 on the same footing. This allows us to track the dynamics of both atomic states and photons, to identify and  classify all the radiative processes, and 
 to compute cross-sections and radiative power without appealing to semiclassical expressions of the sort of that for $\mathcal{W}_{coh}$. 
 We find that, beside a general attenuation of the spectral amplitudes, the pump reduces the nonradiative losses and provides 
gains in the form of stimulated emission and incoherent radiation. In turn, for sufficiently strong pump, gains and losses compensate, resulting in the vanishing of extinction, one of the necessary 
conditions for parity-time-reversal ($\mathcal{PT}$) symmetry in an optical system \cite{Bender,ManjavacasPT}.

The latter finding is intended to pave the way for the extension of the present work to  many-atom systems with gains and losses \cite{Hang,Japs}. In particular, 
we will be ultimately interested in engineering  the exceptional properties already displayed by some semiclassical optical systems \cite{ORN19}. 
In a subsequent work we will address the optical response of a pair of identical atoms, with one of them continuously pumped, and study the anomalies already 
found in analogous systems 
related to $\mathcal{PT}$-symmetry \cite{ManjavacasPT,Khandekar,JKP16}.

The article is organized as follows. In Sec.~\ref{lasec2} we describe the fundamentals of the approach. In Sec.~\ref{lasec3} we identify diagrammatically all the 
 radiative processes and compute the corresponding cross-sections and emitted power.  In Sec.~\ref{subplot} we discuss the energetic balance and  represent graphically the cross-sections in terms of the pump rate. In Sec.~\ref{lasec5} we verify that our results are compatible with unitarity. In passing, we compare our results with those of a semiclassical calculation. 
 The conclusions are summarized in Sec.~\ref{lasec6}.

\begin{figure}
\begin{center}
\includegraphics[width=70mm,angle=0,clip]{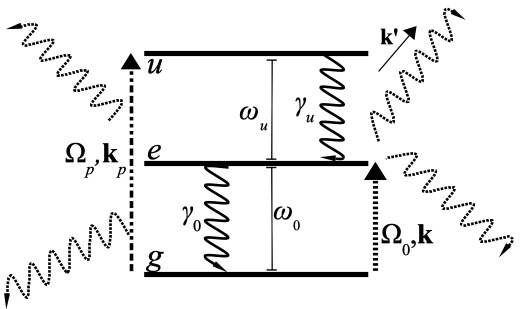}
\caption{Schematics of the system under study, consisting of a three-level atom and two external fields. A pump field of strength $\Omega_{p}$ causes the transient 
excitation of the atom from the ground state $g$ to the upper level $u$, from which it decays, rapidly and incoherently, to the intermediate level $e$ at a rate 
$\gamma_{u}$, causing an effective pump rate $\mathcal{P}=\Omega_{p}^{2}/\gamma_{u}$. Energy intervals and dissipative channels are depicted. 
The atom is illuminated by a weak probe field of strength 
$\Omega_{0}$ and momentum \textbf{k}, quasi-resonant 
with the $g\rightarrow e$ transition. Radiation of momentum $\bf{k'}$ is scattered.} \label{fig1}
\end{center}
\end{figure}

\section{Fundamentals of the approach}\label{lasec2}

Let us consider a three-level atom, with $g$, $e$, and $u$ being the ground state, the excited state, and the upper auxiliary state, respectively, with energy 
intervals $\hbar\omega_{0}$ between $g$ and $e$, and $\hbar\omega_{u}$ between $e$ and $u$, and natural line widths $\gamma_{0}$ and $\gamma_{u}$, 
respectively --see Fig.\ref{fig1}. The atom is continuously illuminated by two linearly polarized 
monochromatic fields referred to as pump and probe fields,  of amplitudes, frequencies, momenta and polarization vectors $E_{0}$, $\omega$, $\mathbf{k}$, $\bm{\epsilon}$, 
and $E_{p}$, $\omega_{p}$, $\mathbf{k}_{p}$, $\bm{\epsilon}_{p}$, respectively. For the sake of simplicity we consider the probe field quasiresonant with the $g\rightarrow e$ 
transition, $|\omega-\omega_{0}|\ll\gamma_{0}$; and the pump field resonant with the $g\rightarrow u$ transition, $\omega_{p}\approx\omega_{0}+\omega_{u}$. Their corresponding 
 Rabi frequencies are $\Omega_{0}=E_{0}\boldsymbol{\mu}\cdot\bm{\epsilon}$ and 
$\Omega_{p}=E_{p}\tilde{\boldsymbol{\mu}}\cdot\bm{\epsilon}_{p}$, with $\boldsymbol{\mu}=\langle g|\mathbf{d}|e\rangle$ and 
$\tilde{\boldsymbol{\mu}}=\langle g|\mathbf{d}|u\rangle$ being the dipole transition moments, and $\mathbf{d}$ being the electric dipole operator.  
Incoherent pumping is achieved for $\gamma_{u}\gg\Omega_{p},\gamma_{0}$, in which case the fast dynamics of the auxiliary state can be integrated out in an effective 
manner. Further, the probe field interacts weakly with the effective two-level atom for $\Omega_{0}\ll\gamma_{0}$. In the following we describe our Hamiltonian 
approach, we quantify the incoherent processes that lead to the reduction of the three-level dynamics to that of an effective two-level atom, and explain how to 
account for the quantum interaction of the probe field with the resultant two-level atom. 

\subsection{Hamiltonian approach}
Our wavefunction-diagrammatic approach is based on the time propagator of the atom-EM field system, $\mathbb{U}(t)$. In terms of the 
Hamiltonian of the system,  $H$ , it reads 
\begin{equation}
\mathbb{U}(t-t_{0})=\mathcal{T}\textrm{-exp}\left\{-i\hbar^{-1}\int_{t_{0}}^{t}d\tau\:H(\tau)\right\},
\end{equation}
where $H$ contains a free component, $H_{0}$, and an interaction term, $W$. As for the free Hamiltonian it reads
\begin{align}
H_{0}&=\hbar\omega_{0}\ket{e}\bra{e}+\hbar(\omega_{0}+\omega_{u})\ket{u}\bra{u}\nonumber\\
&+\sum_{\mathbf{k}',\bm{\epsilon}'}\hbar\omega'(a^{\dagger}_{\mathbf{k}',\bm{\epsilon}'}a_{\mathbf{k}',\bm{\epsilon}'}+\frac{1}{2}),\nonumber 
\end{align}
where the second term corresponds to the free EM Hamiltonian, with $a_{\mathbf{k}',\bm{\epsilon}'}^{\dagger}$ and 
$a_{\mathbf{k}',\bm{\epsilon}'}$ being the creation and annihilation operators of photons of momentum  $\mathbf{k}'$, frequency $\omega'=c\:k'$ and polarization vector $\bm{\epsilon}'$. 
The atom-field interaction is, in the electric dipole approximation,
\begin{equation}
W=-\mathbf{d}\cdot\mathbf{E}(\mathbf{r}_{A}),\nonumber
\end{equation}
where $\mathbf{r}_{A}$ is the atomic center of mass. In Schr\"odinger's picture, 
the electric field operator can be expanded as a sum over normal modes,
\begin{align}
\mathbf{E}(\mathbf{r})&=i\sum_{\mathbf{k}',\bm{\epsilon}'}\sqrt{\frac{\hbar\omega'}{2\epsilon_{0}\mathcal{V}}}
\left[\bm{\epsilon}'a_{\mathbf{k}',\bm{\epsilon}'}e^{i\mathbf{k}'\cdot\mathbf{r}}-\bm{\epsilon}'^{\ast}a^{\dagger}_{\mathbf{k}',\bm{\epsilon}'}
e^{-i\mathbf{k}'\cdot\mathbf{r}}\right]\nonumber\\
&=\sum_{\mathbf{k}',\bm{\epsilon}'}\left[\mathbf{E}^{(+)}_{\mathbf{k}',\bm{\epsilon}'}(\mathbf{r})+\mathbf{E}^{(-)}_{\mathbf{k}',\bm{\epsilon}'}
(\mathbf{r})\right].\label{fieldE}
\end{align}
Essential in our calculations is the vacuum expectation value of the quadratic fluctuations of the electric field, which reads
\begin{equation}
\sum_{\bm{\epsilon}'}\int_{0}^{4\pi}\frac{d\Theta_{\mathbf{k}'}}{8\pi^{2}}\bra{0}\mathbf{E}^{(+)}_{\mathbf{k}',\bm{\epsilon}'}(\mathbf{r})
\mathbf{E}^{(-)}_{\mathbf{k}',\bm{\epsilon}'}(\mathbf{r}')\ket{0}=\frac{-\hbar c}{\epsilon_{0}}
\textrm{Im}\mathbb{G}(\mathbf{r}-\mathbf{r}';\omega').\nonumber
\end{equation}
Here,  $\mathbb{G} (\mathbf{r}-\mathbf{r}';\omega')$ is the dyadic Green's function of the electric field induced at $\mathbf{r}$ by an electric dipole of 
frequency $\omega'$ located at $\mathbf{r}'$, 
\begin{equation}
\mathbb{G} (\mathbf{R};\omega')=-\frac{k' e^{ik'R}}{4\pi}\left[\frac{\mathbb{P}}{k'R}+\frac{i\mathbb{Q}}{(k'R)^{2}}
-\frac{\mathbb{Q}}{(k'R)^{3}}\right],\label{Green}
\end{equation}
where the tensors $\mathbb{P}$ and $\mathbb{Q}$ read $\mathbb{P}=\mathbb{I}-\mathbf{R}\mathbf{R}/R^{2}$,  $\mathbb{Q}=\mathbb{I}-3\mathbf{R}\mathbf{R}/R^{2}$, 
with $\mathbf{R}=\mathbf{r}-\mathbf{r}'$, $k'=\omega'/c$.

Considering $W$ as a perturbation to $H_{0}$, the time propagator of the system admits an expansion in powers of  $W$ which can be developed from its 
time-ordered exponential expression,
\begin{equation}
\mathbb{U}(t-t_{0})=\mathbb{U}_{0}(t)\:\textrm{T-exp}\int_{t_{0}}^{t} -i\hbar^{-1}\mathop{d\tau} \mathbb{U}_{0}^{\dagger}(\tau)W
\mathbb{U}_{0}(\tau-t_{0}),\label{propagator}
\end{equation}
where $\mathbb{U}_{0}(t-t')$ is the unperturbed time-propagator, $\mathbb{U}_{0}(t-t')=\exp{[-i\:H_{0}(t-t')]}$.

\subsection{Incoherent dynamics with the pump field. Effective two-level atom}\label{incoherent}
\begin{figure}
\begin{center}
\includegraphics[height=95mm,width=90mm,angle=0,clip]{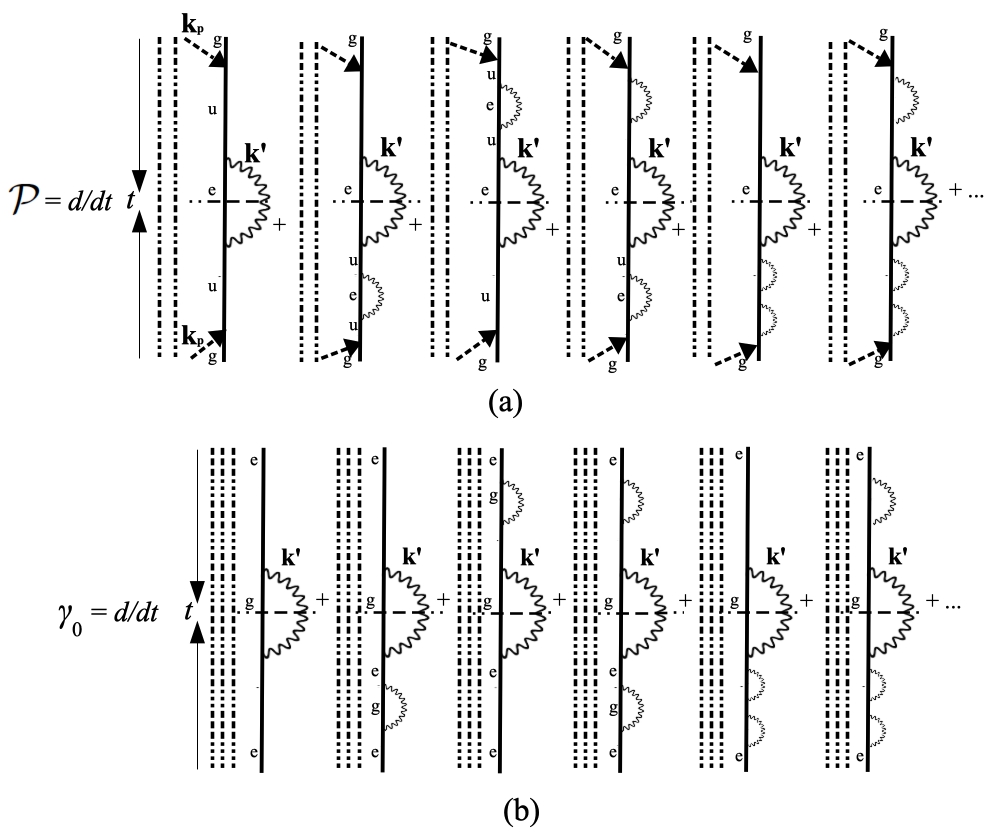}
\caption{Diagrammatic representation of the processes that contribute to the pump rate, $\mathcal{P}$, and to the spontaneous decay rate, $\gamma_{0}$. 
Time runs along the vertical axis towards the observation time, $t$. Photons of the pump field with momentum $\mathbf{k}_{p}$ are depicted with straight dashed-dotted arrows --vertical dashed-dotted lines stand for noninteracting-spectator photons , while emitted photons of undefined momentum $\mathbf{k}'$ appear as wavy lines.}\label{fig2p}
\end{center}
\end{figure}

Let us consider first the action of the pump field alone under the condition $\gamma_{u}\gg\Omega_{p},\gamma_{0}$. 
The  corresponding quantum state of the EM field is denoted by
\begin{equation}
|N_{\mathbf{k}_{p},\bm{\epsilon}_{p}}\rangle=\frac{1}{\sqrt{N_{\mathbf{k}_{p},\bm{\epsilon}_{p}}!}}
\left(a_{\mathbf{k}_{p},\bm{\epsilon}_{p}}^{\dagger}\right)^{N_{\mathbf{k}_{p},\bm{\epsilon}_{p}}}\ket{0},\nonumber
\end{equation}
where $N_{\mathbf{k}_{p},\bm{\epsilon}_{p}}/\mathcal{V}=\epsilon_{0}E_{p}^{2}/\hbar\omega_{p}$ is the pump-photon density,  with $\mathcal{V}$ being a quantization volume,
and $|0\rangle$ is the EM vacuum state. The incoherent dynamics is determined by the rapid decay of the atom from the state $u$ to $e$ after the action 
of the pump upon the atom in state $g$--Fig.\ref{fig2p}(a), and the 'slow' decay from the state $e$ to $g$ --Fig.\ref{fig2p}(b). Considering these phenomena as 
Markovian, they are the result  
of a series of consecutive processes of emission and reabsorption of single photons. The addition of all the diagrams in Fig.\ref{fig2p}(a) yields for the 
rate of the incoherent pump transition, $\mathcal{P}=d|\langle e|\mathbb{U}(t)|g\rangle|^{2}/dt$, 
\begin{align}
\mathcal{P}&=-2\frac{\Omega_{p}^{2}}{\gamma_{u}^{2}}\frac{(\omega_{u}-\omega_{0})^{2}}{\hbar\epsilon_{0}c^{2}}\hat{\boldsymbol{\mu}}\cdot\textrm{Im}\:\mathbb{G}(\mathbf{R};\omega_{u}-\omega_{0})\cdot\hat{\boldsymbol{\mu}}\nonumber\\
&=\frac{\Omega_{p}^{2}}{\gamma_{u}},\quad R\rightarrow0^{+},\label{eqP}
\end{align}
where $\hat{\boldsymbol{\mu}}=\langle u|\mathbf{d}|e\rangle$ and, in the last equality we identify 
\begin{equation}
\gamma_{u}=-\frac{2(\omega_{u}-\omega_{0})^{2}}{\hbar\epsilon_{0}c^{2}}\hat{\boldsymbol{\mu}}\cdot\textrm{Im}\mathbb{G}(\mathbf{R};\omega_{u}-\omega_{0})\cdot\hat{\boldsymbol{\mu}},\:R\rightarrow0^{+}.\nonumber
\end{equation}
The reading of Fig.\ref{fig2p}(a) in terms of quantum states and operators is compiled in the Appendix \ref{app1}. For the sake of 
completeness, the rate of spontaneous decay from $e$ to $g$ from the diagrams in Fig.\ref{fig2p}(b) is
\begin{equation}
\gamma_{0}=\frac{-2\omega_{0}^{2}}{\hbar\epsilon_{0}c^{2}}\boldsymbol{\mu}\cdot\textrm{Im}\:\mathbb{G}(\mathbf{R};\omega_{0})\cdot\boldsymbol{\mu},\quad R\rightarrow0^{+}.\label{eqg}
\end{equation}
For $\gamma_{u}\gg\Omega_{p},\gamma_{0}$, the dynamics of the state $u$ can be integrated out adiabatically. In terms of the usual nomenclature of the density 
functional formalism, this implies the approximation $\mathcal{P}\rho_{gg}\approx\gamma_{u}\rho_{uu}$, which leads to the following Bloch's equations 
for the effective two-level atom, 
\begin{align}
\partial_{t}\rho_{ee}&=-\gamma\rho_{ee}+\mathcal{P}\rho_{gg},\label{2leveleqsa}\\
\partial_{t}\rho_{gg}&=\gamma\rho_{ee}-\mathcal{P}\rho_{gg},\label{2leveleqsb}\\
\partial_{t}\rho_{eg}&=-(\gamma+\mathcal{P})\rho_{eg}/2-i\omega_{0}\rho_{eg}.\label{2leveleqsc}
\end{align}
In these equations we have allowed for nonradiative contributions to the decay rate from $e$ to $g$, $\gamma_{nr}$, by introducing $\gamma=\gamma_{nr}+\gamma_{0}$ in the place 
of $\gamma_{0}$. Let us note that these equations take account of the effective dynamics of the atomic degrees of freedom only, which are the ones of interest in 
the incoherent dynamics. Equivalent equations are obtained starting with Bloch's equations for the three states \cite{German,Lagendijk_3_level,TejedorII}. 
When dealing with the photonic degrees of freedom in our wavefunction formalism, these equations will provide the attenuating factors associated to the incoherent 
processes. From Eq.(\ref{2leveleqsc}) for the coherence element $\rho_{eg}$ we read that the pump attenuates the coherent evolution between the two levels 
\cite{TejedorII}. Defining $\Gamma=\gamma+\mathcal{P}$, straightforward integration of the above equations leads to the solutions
\begin{align}\label{populations}
\rho_{ee}(t)&=\frac{\mathcal{P}}{\Gamma}(1-e^{-\Gamma t})+N_{e}e^{-\Gamma t},\\
\rho_{gg}(t)&=\frac{\gamma}{\Gamma}(1-e^{-\Gamma t})+e^{-\Gamma t}(1-N_{e}),\\
\rho_{eg}(t)&=\sqrt{N_{e}-N_{e}^{2}}\:e^{-i\omega_{0}t}e^{-\Gamma t/2},
\end{align}
where $N_{e}$ is the population of the excited state at $t=0$. For asymptotic times, $\Gamma t\gg1$, the atomic populations converge to the 
steady values $\rho_{gg}(t\rightarrow\infty)=\gamma/\Gamma$, $\rho_{ee}(t\rightarrow\infty)=\mathcal{P}/\Gamma$, 
while the coherence element vanishes at a rate $\Gamma$ regardless of the initial conditions.

\subsection{Wavefunction approach with the probe field}
Having parametrized the effecive action of the 
pump field with the pump rate $\mathcal{P}$, we omit its contribution to the quantum EM state hereafter and consider the contribution of the probe field photons only,
\begin{equation}
|N_{\mathbf{k},\bm{\epsilon}}\rangle=\frac{1}{\sqrt{N_{\mathbf{k},\bm{\epsilon}}!}}
\left(a_{\mathbf{k},\bm{\epsilon}}^{\dagger}\right)^{N_{\mathbf{k},\bm{\epsilon}}}\ket{0},\nonumber
\end{equation}
where $N_{\mathbf{k},\bm{\epsilon}}/\mathcal{V}=\epsilon_{0}E_{0}^{2}/\hbar\omega$ is the photon density and  $\epsilon_{0}c\!E_{0}^{2}/2$ is the time-averaged intensity.
Therefore, once the atomic state has reached its steady state, the atom-EM field state is a mixed state made of the incoherent superposition of the pure states
\begin{equation}
|\Psi_{0}\rangle_{g}=\sqrt{\gamma/\Gamma}|N_{\mathbf{k},\bm{\epsilon}};g\rangle,\quad
|\Psi_{0}\rangle_{e}=\sqrt{\mathcal{P}/\Gamma}|N_{\mathbf{k},\bm{\epsilon}};e\rangle,
\label{wavefunction}
\end{equation}
where we have included the statistical weights, $\sqrt{\gamma/\Gamma}$ and $\sqrt{\mathcal{P}/\Gamma}$, respectively. 
It is upon the statistical mixture of these states that quantum perturbation theory is to be applied in the computation of the optical response. Thus, the physical 
quantities to be calculated are statistical averages over the quantum expectation values computed upon the pure states $|\Psi_{0}\rangle_{g}$ and 
$|\Psi_{0}\rangle_{e}$.

\begin{figure}
\begin{center}
\includegraphics[width=85mm,angle=0,clip]{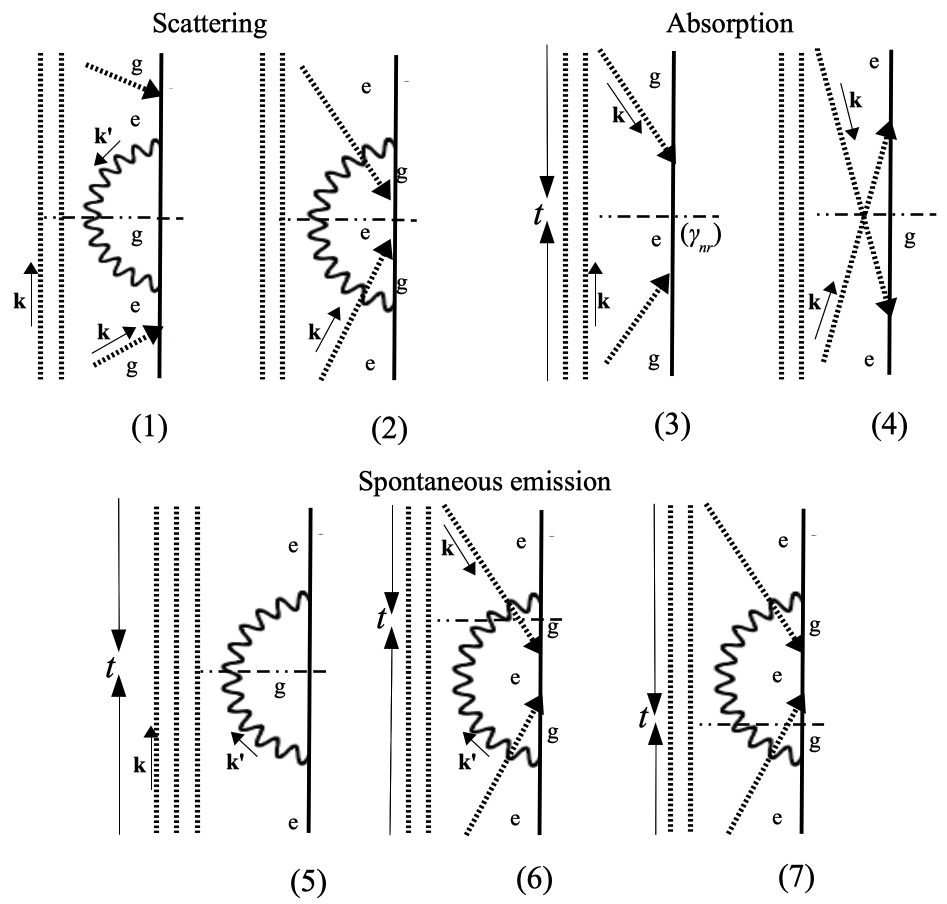}
\caption{Diagrammatic representation of the seven processes that contribute, at leading order, to the single-atom cross-sections of scattering [(1) and (2)] and absorption [(3) and (4)], as well as to  
spontaneous emission [(5)-(7)].  The symbol $(\gamma_{nr})$ in diagram (3) indicates that only nonradiative dissipation is to be accounted in that process. 
Time runs along the vertical axis towards the observation time, $t$. Photons of the incident probe field with momentum $\mathbf{k}$ are depicted with straight 
dashed arrows --vertical dashed lines stand for noninteracting-spectator photons , while emitted photons of undefined momentum $\mathbf{k}'$ appear as wavy lines.  
The incoherent superposition of the states $|\Psi_{0}\rangle_{g}$ and $|\Psi_{0}\rangle_{e}$ implies the absence of interference between the 
wavefunctions of both states.}\label{fig2}
\end{center}
\end{figure}

For a weak probe field, $\Omega_{0}\ll\Gamma$, the optical response of the atom at leading order involves terms of up to $\mathcal{O}(W^{4})$ in 
$\mathbb{U}$ in Eq.(\ref{propagator}).  They are represented diagrammatically in Fig.\ref{fig2}. 
In all of them, except for diagram (5), two of the interaction vertices, $W$, create or annihilate one photon of the probe field each.

Finally, from Eq.(\ref{2leveleqsc})  
we read that the coherent transitions from steady to intermediate states, say, from 
$|\Psi_{0}\rangle_{g}$ to $|N_{\mathbf{k},\bm{\epsilon}},1_{\mathbf{k}',\bm{\epsilon}'};e\rangle$ or from $|\Psi_{0}\rangle_{e}$ to 
$|N_{\mathbf{k},\bm{\epsilon}},1_{\mathbf{k}',\bm{\epsilon}'};g\rangle$ in a time interval $t$, get attenuated in time at an effective rate 
$\Gamma/2$. That is, at leading order in $W$, using Eq.(\ref{propagator}), 
$\langle N_{\mathbf{k},\bm{\epsilon}},1_{\mathbf{k}',\bm{\epsilon}'};e|\mathbb{U}(t)|\Psi_{0}\rangle_{g}\sim
\sqrt{\gamma/\Gamma}\int_{0}^{t}d\tau\:e^{-\Gamma(t-\tau)/2}e^{-i(\omega_{0}+\omega')(t-\tau)}
\langle e|\mathbf{d}|g\rangle$.

\section{Identification of radiative processes. Computation of power and cross-sections}\label{lasec3}

Let us consider the atom in the steady state of Eq.(\ref{wavefunction}), continuously illuminated by the 
 weak and quasi-resonant probe field of strength $\Omega_{0}$ and frequency $\omega$ as outlined in Sec.~\ref{lasec2}. 
 In the following we identify all the radiative and non-radiative processes 
 which contribute, at leading order in $W$, to the scattered power, spontaneous emission, and  cross-sections 
 of extinction, absorption, and scattering. To this end, we expand the time propagator of Eq.(\ref{propagator}) up to terms of order $W^{4}$ and represent 
 diagrammatically the probabilities of the processes $\langle\Psi_{n}^{f}|\Psi_{n}(t)\rangle$, i.e., $P_{n}(t)=|\langle\Psi_{n}^{f}|\Psi_{n}(t)\rangle|^{2}$.  
 In this expression the state $|\Psi_{n}(t)\rangle$ results from the evolution of one of the pure states, $|\Psi_{0}\rangle_{g}$ or $|\Psi_{0}\rangle_{e}$, 
 in a time interval $t$, $|\Psi_{n}(t)\rangle=\mathbb{U}(t)|\Psi_{0}\rangle_{g/e}$; and the state $|\Psi_{n}^{f}\rangle$ is that whose radiative content is 
 to be computed for the calculation of the radiative power. Note that, in all the cases, the radiative content of $|\Psi_{n}^{f}\rangle$ differs 
 from that of the pure states $|\Psi_{0}\rangle_{g/e}$  either in the net number of photons, or in the frequency, momentum and polarization of the photons.
 The corresponding processes are represented in Fig.\ref{fig2} and are labeled with the subscript $n$. Since the illumination is continuous, we will be interested 
 in steady processes for which we consider asymptotic times, $\Gamma t\gg1$.\\
 
\subsection{Scattering}\label{scatt}
 \noindent Scattering involves processes in which  the atomic state  in  $|\Psi_{n}^{f}\rangle$ coincides with that in  $|\Psi_{n}(0)\rangle$, 
 but one of the probe photons in $|\Psi_{n}(0)\rangle$ is replaced in 
 $|\Psi_{n}^{f}\rangle$ with a scattered photon of undefined frequency $\omega'$, momentum $\mathbf{k}'$, and polarization $\bm{\epsilon}'$ upon integration. 
 That corresponds to diagrams (1) and (2)  in Fig.\ref{fig2}, which represent scattering from the pure ground state and from the pure excited state, respectively, such that
\begin{align}
|\Psi_{1}(0)\rangle&=|\Psi_{0}\rangle_{g},\quad|\Psi_{1}^{f}\rangle=\sum_{\mathbf{k}',\bm{\epsilon}'}|(N-1)_{\mathbf{k},\bm{\epsilon}},1_{\mathbf{k}',\bm{\epsilon}'};g\rangle,\nonumber\\
|\Psi_{2}(0)\rangle&=|\Psi_{0}\rangle_{e},\quad|\Psi_{2}^{f}\rangle=\sum_{\mathbf{k}',\bm{\epsilon}'}|(N-1)_{\mathbf{k},\bm{\epsilon}},1_{\mathbf{k}',\bm{\epsilon}'};e\rangle.\nonumber
\end{align}
In the former case the emission of the scattered photon follows the absorption of the probe field photon, whereas in the 
latter emission precedes absorption. Important is to note that, in the steady state and under continuous illumination, $\Gamma t\gg1$, the 
process of absorption of a probe field photon in diagram (1) is followed without delay by the emission of a photon of undefined momentum. Thus, the emission 
process does not correspond to spontaneous emission, but to continuous scattering \cite{Sakurai}. Likewise, under 
continuous illumination, in diagram (2) the emission of a photon of undefined momentum is followed without delay by the absorption of a probe field photon, 
resulting in continuous scattering. Things would be different for the case that the probe field were a short pulse, in which case absorption would be followed by 
spontaneous emission. Therefore,  in scattering processes the transition between atomic states is just transient and the scattered power is the time derivative 
of the EM energy, 
\begin{align}
\mathcal{W}_{sc}=&\sum_{n=1}^{2}\frac{\mathop{d}}{\mathop{dt}}\langle\Psi_{n}(t)|\Psi_{n}^{f}\rangle\langle\Psi_{n}^{f}|H_{EM}|\Psi_{n}^{f}\rangle
\langle\Psi_{n}^{f}|\Psi_{n}(t)\rangle\nonumber\\
=&\frac{\hbar\omega\Omega_{0}^{2}\gamma_{\omega}}{4[(\omega-\omega_{0})^{2}+\Gamma^{2}/4]},\quad\Gamma t\gtrsim1, \label{powerscat} 
\end{align}
where $\gamma_{\omega}=\frac{-2\omega^{2}}{c^{2}\epsilon_{0}\hbar}\boldsymbol{\mu}\cdot\textrm{Im}\:\mathbb{G}(\mathbf{R};\omega)\cdot\boldsymbol{\mu}$, $R\rightarrow0^{+}$, 
and the scattering cross-section reads
\begin{equation}
\sigma_{sc}= \frac{2\hbar\omega}{c\epsilon_{0}E_{0}^{2}}\frac{\mathop{d\left(P_{1}+P_{2}\right)}}{\mathop{dt}}=
\frac{\mu_{\parallel}^{2}\omega\gamma_{\omega}/(2\hbar\epsilon_{0}c)}{(\omega-\omega_{0})^{2}+\Gamma^{2}/4}, \label{sigmascat} 
\end{equation}
where $\mu_{\parallel}=\boldsymbol{\mu}\cdot\bm{\epsilon}$. The reading of the contribution of diagrams (1) and (2) to $\mathcal{W}_{sc}$ in terms of 
quantum operators and states is compiled in Appendix \ref{app2}.
 
\subsection{Absorption, stimulated emission and extinction}
\noindent Generically, absorption processes are those in which  states  $|\Psi_{n}(0)\rangle$ 
and $|\Psi_{n}^{f}\rangle$ differ both in the atomic states and in the number of probe field photons. 
They are represented by diagrams (3) and (4) of Fig.\ref{fig2}. 
In diagram (3) the atom at time $t$ gets excited with respect to its initial state, while the EM state 
contains one probe photon less than the initial state. In diagram (4) the atom gets de-excited at time $t$ while the EM state contains one probe photon 
more than the initial state,
\begin{align}
|\Psi_{3}(0)\rangle&=|\Psi_{0}\rangle_{g},\quad|\Psi_{3}^{f}\rangle=|(N-1)_{\mathbf{k},\bm{\epsilon}};e\rangle,\nonumber\\
|\Psi_{4}(0)\rangle&=|\Psi_{0}\rangle_{e},\quad|\Psi_{4}^{f}\rangle=|(N+1)_{\mathbf{k},\bm{\epsilon}};g\rangle.\nonumber
\end{align}
The former process contributes to positive absorption, whereas the latter accounts for stimulated emission or negative absorption. In the steady state, 
under continuous  illumination,  the transition $g\leftrightarrow e$  induced by the potential $W$ 
in either direction takes place at a constant rate $\Gamma$, as that is the coherence time sets by the pump. Thus, the rate at which an absorptive process takes 
place is $\Gamma$ times  the probability that such a process takes 
 place for asymptotic times $\Gamma t\gg1$. Also, since the states $|\Psi_{0}\rangle_{g}$ and  $|\Psi_{0}\rangle_{e}$ are stationary, absorption is necessarily  
 followed  by processes that take the  
 atomic state at time $t$ back to the atomic state at time $0$. In particular, the state $e$ decays  into $g$  
 either emitting a scattered photon of frequency $\omega$ 
 according to diagram (1) in Fig.\ref{fig2}, or by nonraditive means. Therefore, in order not to double-count the probability of radiative decay, the contribution of 
 diagram (1) must be substracted from that of diagram (3) in the calculation of net absorption, resulting in nonradiative absorption only. 
 Likewise, the state $g$ at time $t$ in diagram (4) 
 ends up transiting to state $e$ under the action of the pump which, according to diagram $(a)$ in Fig.\ref{fig2p}, is accompanied by the spontaneous emission of 
 photons of frequency $\simeq\omega_{u}-\omega_{0}$. Since no other radiative processes are involved there, 
 no double-counting is associated to the probability of the process in diagram (4).
 
 Thus, the power absorbed by the system is written as 
 \begin{align}
\mathcal{W}_{abs}&=\Gamma\sum_{n=3}^{4}\Bigl[\langle\Psi_{n}(0)|H_{EM}|\Psi_{n}(0)\rangle
-\langle\Psi_{n}(t)|\Psi_{n}^{f}\rangle\nonumber\\
&\times\langle\Psi_{n}^{f}|H_{EM}|\Psi_{n}^{f}\rangle\langle\Psi_{n}^{f}|\Psi_{n}(t)\rangle\Bigr]
-\frac{\mathop{d}}{\mathop{dt}}\langle\Psi_{1}(t)|\Psi_{1}^{f}\rangle\nonumber\\
&\times\langle\Psi_{1}^{f}|H_{EM}|\Psi_{1}^{f}\rangle
\langle\Psi_{1}^{f}|\Psi_{1}(t)\rangle=\frac{\hbar\omega\Omega_{0}^{2}\left(\gamma-\mathcal{P}-\frac{\gamma}{\Gamma}\gamma_{\omega}\right)}{4[(\omega-\omega_{0})^{2}+\Gamma^{2}/4]}\nonumber\\
&=\frac{\hbar\omega\Omega_{0}^{2}\left(\frac{\gamma}{\Gamma}\gamma_{nr}-\frac{\mathcal{P}}{\Gamma}\mathcal{P}\right)}{4[(\omega-\omega_{0})^{2}+\Gamma^{2}/4]},\quad
\Gamma t\gtrsim1.\label{powerabs} 
\end{align}
where the time-derivative term with a minus sign in front stems from the subtraction of the scattered power from the pure ground state. As for the absorption-cross section, 
\begin{align}
\sigma_{abs}&=\frac{2\hbar\omega}{c\epsilon_{0}E_{0}^{2}}\left[\Gamma(P_{3}-P_{4})-\frac{\mathop{d\:P_{1}}}{\mathop{dt}}\right]\nonumber\\
&\simeq\frac{\mu_{\parallel}^{2}\omega\left(\frac{\gamma}{\Gamma}\gamma_{nr}-\frac{\mathcal{P}}{\Gamma}\mathcal{P}\right)/(2\hbar\epsilon_{0}c)}
{(\omega-\omega_{0})^{2}+\Gamma^{2}/4},\label{sigmabs} 
\end{align}
where the minus sign in front of $P_{4}$ accounts for the negative nature of the absorption associated to stimulated emission. 
The reading of the contributions of diagrams (3) and (4) to $\mathcal{W}_{abs}$ are compiled in Appendix \ref{app2}.
 

Finally, the extinction cross-section is the addition of $\sigma_{sc}$ and $\sigma_{abs}$, which yields
\begin{align}
\sigma_{ext}&\simeq\frac{\mu_{\parallel}^{2}\omega\left(\gamma_{\omega}+\frac{\gamma}{\Gamma}\gamma_{nr}-\frac{\mathcal{P}}{\Gamma}\mathcal{P}\right)/(2\hbar\epsilon_{0}c)}
{(\omega-\omega_{0})^{2}+\Gamma^{2}/4}\nonumber\\
&=\frac{\mu_{\parallel}^{2}\omega\left[\gamma-\frac{\mathcal{P}}{\Gamma}(\gamma_{nr}+\mathcal{P})\right]/(2\hbar\epsilon_{0}c)}{(\omega-\omega_{0})^{2}+\Gamma^{2}/4}.\label{sigmaext}
\end{align}

\subsection{Spontaneous emission}
Lastly, spontaneous emission, up to terms of order $\Omega_{0}^{2}/\Gamma^{2}$, corresponds to the processes depicted by diagrams 
(5), (6) and (7) in Fig.\ref{fig2}. In all of them the atom transits from the excited state at time $0$ to the ground state at time $t$, and $|\Psi_{n}^{f}\rangle$ 
contains the same number of probe field photons as 
$|\Psi_{n}(0)\rangle$, plus one more photon of undefined frequency, momentum and polarization upon integration,
\begin{equation}
|\Psi_{5,6,7}(0)\rangle=|\Psi_{0}\rangle_{e},\quad|\Psi_{5,6,7}^{f}\rangle=\sum_{\mathbf{k}',\bm{\epsilon}'}|N_{\mathbf{k},\bm{\epsilon}},1_{\mathbf{k}',\bm{\epsilon}'};g\rangle.\nonumber
\end{equation}
Diagram (5) is similar to those for the natural decay rate in the absence of the probe field in Fig.\ref{fig2p}(b), $\gamma_{0}$, 
but for the fact that the pump enhances the effective incoherence rate and thus the width of the emission line. 
On the other hand, diagrams (6) and (7) depict the influence of the probe field on spontaneous emission. Note that, in contrast to the scattering diagram (2), a probe field 
photon in the final state is not effectively absorbed, as it reappears in the final state. As with absorption,  under steady conditions, the rate at which the 
spontaneous emission processes take place is $\Gamma$ times their probabilities for asymptotic times, $\Gamma t\gg1$. Spontaneous emission 
is inherently incoherent (\emph{inc}), and its power reads   
\begin{align}
\mathcal{W}_{inc}&=\Gamma\sum_{n=5}^{7}\Bigl[\langle\Psi_{n}(t)|\Psi_{n}^{f}\rangle\langle\Psi_{n}^{f}|H_{EM}|\Psi_{n}^{f}\rangle
\langle\Psi_{n}^{f}|\Psi_{n}(t)\rangle\nonumber\\
&-\langle\Psi_{n}(0)|H_{EM}|\Psi_{n}(0)\rangle\Bigr]
=\frac{\mathcal{P}}{\Gamma}\Bigl\{\hbar\omega_{0}\gamma_{0}\nonumber\\
&-\frac{\Omega_{0}^{2}\Gamma^{2}/8}{[(\omega-\omega_{0})^{2}+\Gamma^{2}/4]^{2}}\Bigl[\hbar\omega\gamma_{\omega}-\hbar\omega_{0}\gamma_{0}/2\nonumber\\
&+2\hbar\omega_{0}\gamma_{0}(\omega-\omega_{0})^{2}/\Gamma^2\Bigr]\Bigr\}\label{powerincoh}\\
&\simeq\frac{\hbar\omega_{0}\gamma_{0}\mathcal{P}}{\Gamma}\Bigl[1-\frac{\Omega_{0}^{2}\Gamma^{2}/16}
{[(\omega-\omega_{0})^{2}+\Gamma^{2}/4]^{2}}\Bigr],\quad\Gamma t\gtrsim1.\nonumber
\end{align}
Note that the quasi-resonant probe field in diagrams (6) and (7) generates resonances at $\omega$ in addition to those at $\omega_{0}$ in the spectrum of spontaneous emission. Further, in the limit $|\omega-\omega_{0}|/\Gamma\rightarrow0$, we obtain a probe-field corrected spontaneous emission rate 
$\mathcal{P}\gamma_{0}(1-\Omega_{0}^{2}/\Gamma^{2})/\Gamma$. The reading of the contributions of diagrams (5), (6), and (7) to $\mathcal{W}_{inc}$ are compiled in Appendix \ref{app2}.



\section{Energetics: balance between gain and loss}\label{subplot}

Regarding the energetic content of the radiative processes, the interpretation is as follows. From the expression on the 
right hand side of the first equality in 
Eq.(\ref{sigmaext}) for the extinction cross-section we read that, in addition to the scattering term $\propto\gamma_{\omega}$,  the nonradiative dissipative term 
 $\propto\gamma\gamma_{nr}/\Gamma$, proportional to the population rate of the ground state, accounts for positive losses; and the negative term  
 $\propto-\mathcal{P}^{2}/\Gamma$, proportional to the population rate of the excited state, stems 
 from the stimulated emission fed by the pump. On the other hand, from the expression on the right hand side of the second equality in Eq.(\ref{sigmaext}), 
we read that the net action of the pump on the probe field is that of reducing its extinction in an amount proportional to 
$-\mathcal{P}(\mathcal{P}+\gamma_{nr})/\Gamma$. That is, the excited state population contributes positively to probe field radiation through stimulated emission, 
and reduces the nonradiative losses associated to the transition $e\rightarrow g$ after the transient excitation caused 
by the probe field from the state $g$ to $e$. On top of that, the incoherent power of Eq.(\ref{powerincoh}) is proportional to the population rate of the excited 
state too. In summary, Eq.(\ref{sigmaext}) for $\sigma_{ext}$ differs from that of an atom in its ground state not only in the enhancement of its linewidth and the attenuation of its 
spectral amplitude, but also in the diminishing of nonradiative losses and in the gain provided by stimulated emission. Both effects are proportional to the  
population of the excited state, being the associated energy supplied by the pump. Finally, the incoherent power of Eq.(\ref{powerincoh}) associated to the 
spontaneous decay from the excited to the ground state is supplied by the pump too.
\begin{figure}
\begin{center}
\includegraphics[width=87mm,angle=0,clip]{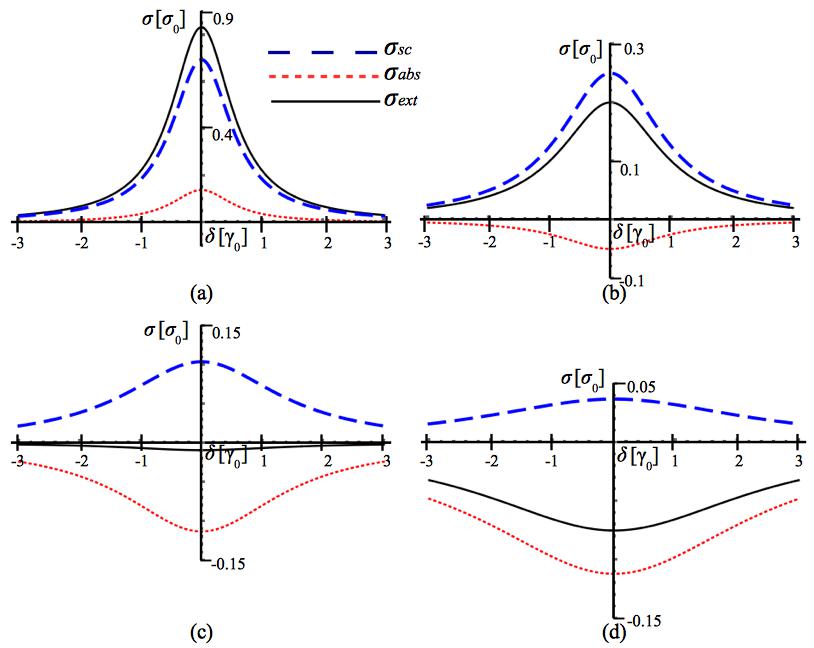}
\caption{Graphical representation of the scattering, absorption and extinction cross-sections for a fixed value of the nonradiative decay rate, $\gamma_{nr}=\gamma_{0}/5$, 
and different values of the pump rate, $\mathcal{P}=0$ [(a)], $\mathcal{P}=0.8\gamma_{0}$ [(b)], $\mathcal{P}=1.9\gamma_{0}$ [(c)], $\mathcal{P}=4\gamma_{0}$ [(d)]. 
Cross-sections are expressed in units of $\sigma_{0}=2\omega_{0}\mu_{\parallel}^{2}/(c\epsilon_{0}\hbar\gamma_{0})$. The detuning $\delta=\omega-\omega_{0}$ 
is given in units of $\gamma_{0}$.}
\label{plots1}
\end{center}
\end{figure}
\begin{figure}
\begin{center}
\includegraphics[width=85mm,angle=0,clip]{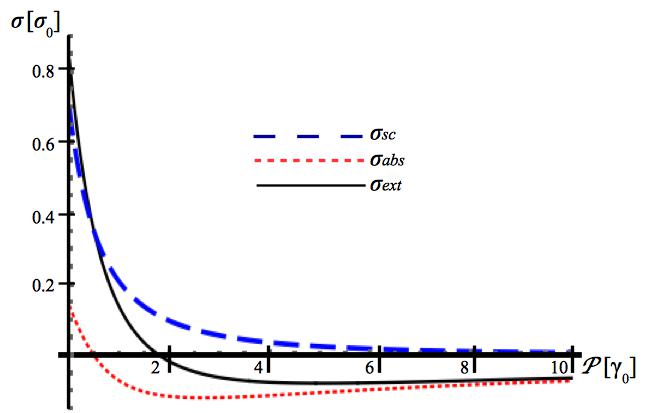}
\caption{Graphical representation of the scattering, absorption and extinction cross-sections for a fixed value of the nonradiative decay rate, 
$\gamma_{nr}=\gamma_{0}/5$, as a function of the pump rate $\mathcal{P}$. Cross-sections are expressed in units of 
$\sigma_{0}=2\omega_{0}\mu_{\parallel}^{2}/(c\epsilon_{0}\hbar\gamma_{0})$. The pump rate is given in units of $\gamma_{0}$.}
\label{plots2}
\end{center}
\end{figure}

We finalize this Section with the graphical representation of the cross-sections  in terms of the parameters of gains and losses, i.e., $\mathcal{P}$ and $\gamma$, 
respectively. From the expressions of Eqs.(\ref{sigmascat}), (\ref{sigmabs}) and (\ref{sigmaext}) for $\sigma_{sc}$, $\sigma_{abs}$ and $\sigma_{ext}$  
we note that the linewidth of all the spectra increases with gains and losses in the same manner,  
$\Gamma=\gamma+\mathcal{P}$, while the spectral amplitudes decrease. However, for the case that either the nonradiative decay rate or the pump rate become dominant,  the scaling behaviors of the cross-sections differ 
from one another.  That is, for  $\gamma_{nr}\gg\gamma_{\omega},\mathcal{P}$,  we find for scattering  $\sigma_{sc}\sim1/\gamma_{nr}^{2}$; while for absorption and extinction we get 
$\sigma_{abs},\sigma_{ext}\sim1/\gamma_{nr}$. Likewise, for $\mathcal{P}\gg\gamma_{nr},\gamma_{\omega}$, we have $\sigma_{sc}\sim1/\mathcal{P}^{2}$ and  
$\sigma_{abs},\sigma_{ext}\sim1/\mathcal{P}$. 
Besides, while scattering is hardly affected by the relationship between gains and losses, absorption and extinction are.  
 In particular, absorption vanishes for $\mathcal{P}=\sqrt{\gamma_{nr}^{2}+\gamma_{nr}\gamma_{\omega}}$, while extinction does 
so for $\mathcal{P}=\left[\gamma_{\omega}+\sqrt{4\gamma_{nr}^{2}+8\gamma_{nr}\gamma_{\omega}+5\gamma_{\omega}^{2}}\right]/2$. The latter equality determines the balance 
between gains and losses which, in an effective manner, is a necessary condition for $\mathcal{PT}$-symmetry \cite{Bender,ManjavacasPT,JKP16}. 
In particular, for $\gamma_{nr}\ll\gamma_{\omega}\simeq\gamma_{0}$, null extinction holds for $\mathcal{P}\simeq\gamma_{0}(1+\sqrt{5})/2$.

In Fig.\ref{plots1} we represent the scattering, absorption and extinction cross-sections for different values of the pump rate. The cross-sections are expressed in units of 
$\sigma_{sc}(\omega=\omega_{0},\Gamma=\gamma_{0})\equiv\sigma_{0}=2\omega_{0}\mu_{\parallel}^{2}/(c\epsilon_{0}\hbar\gamma_{0})$, and the gain and loss rates are given in units 
of $\gamma_{0}$. In Fig.\ref{plots2} we represent the cross-sections at exact resonance, $\omega=\omega_{0}$, in terms of the pump rate at a fixed value of the 
nonradiative decay rate, $\gamma_{nr}=\gamma_{0}/5$. Extinction becomes negative for $\mathcal{P}\gtrsim1.8~\gamma_{0}$.

\section{Discussion}\label{lasec5}

\subsection{Unitarity and Energy balance}

\begin{figure}
\begin{center}
\includegraphics[width=85mm,angle=0,clip]{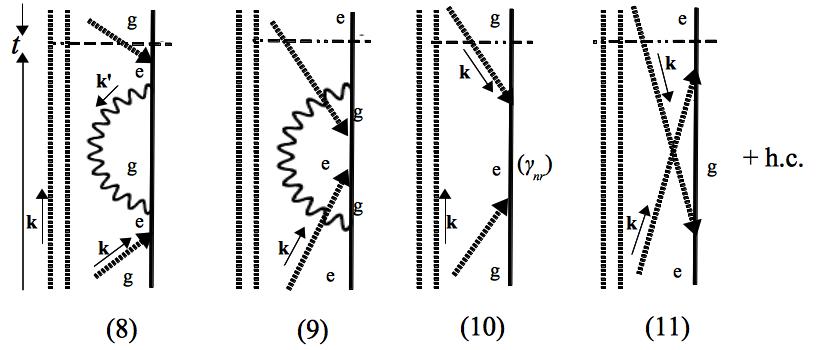}
\caption{Diagrammatic representation of the four processes that, together with their hermitian conjugate versions (h.c.), contribute at leading order to 
$P_{\Psi_{0}\rightarrow\Psi_{0}}$ and thus to the renormalization of the pure steady states $|\Psi_{0}\rangle_{g,e}$.  
As explained in the text, since $|\Psi_{0}\rangle_{g,e}$ are already normalized, the addition of the probabilities of these processes to those of the processes depicted 
in Fig.\ref{fig2} vanishes identically.}\label{fig4}
\end{center}
\end{figure}

On the one hand, our Hamiltonian approach allows us to keep track of the atomic dynamics as well as of all the radiative processes. On the other, 
incoherent processes are accounted for in an effective manner that should be consistent with unitarity. This means that, starting with the normalized mixed steady 
state defined as the incoherent superposition of the pure states in Eq.(\ref{wavefunction}), the addition of the time-derivatives of the probabilities of all the 
processes which take the system to a state different to $|\Psi_{0}\rangle_{g/e}$, $P_{\Psi_{0}\nrightarrow\Psi_{0}}$, and those which take it back to 
$|\Psi_{0}\rangle_{g/e}$, $P_{\Psi_{0}\rightarrow\Psi_{0}}$, 
must be identically zero. This should be the case at all orders. 
In particular, at the order $\Omega^{2}_{0}/\Gamma$, the time-derivative of the probabilities of the radiative processes computed in the previous section amounts to 
\begin{align}
\frac{d}{dt}&P_{\Psi_{0}\nrightarrow\Psi_{0}}|_{\mathcal{O}(\Omega^{2}_{0}/\Gamma)}=
\frac{d}{dt}\left(P_{1}+P_{2}\right)+\Gamma\left(P_{3}+P_{4}+P_{6}+P_{7}\right)\nonumber\\
&\simeq\frac{\Omega_{0}^{2}\left(\gamma+\mathcal{P}+\frac{\mathcal{P}}{\Gamma}\gamma_{\omega}-\frac{\mathcal{P}}{\Gamma}\gamma_{0}\right)}
{4[(\omega-\omega_{0})^{2}+\Gamma^{2}/4]}\simeq\Omega^{2}_{0}/\Gamma,\:\:|\omega-\omega_{0}|\ll\Gamma,\label{Pnpsi}
\end{align}
whereas the processes of Fig.\ref{fig4} give rise to an effective renormalization of the states $|\Psi_{0}\rangle_{g,e}$ proportional to $\Omega_{0}^{2}$,
\begin{align}
\frac{d}{dt}&P_{\Psi_{0}\rightarrow\Psi_{0}}|_{\mathcal{O}(\Omega^{2}_{0}/\Gamma)}=2\frac{d}{dt}\left(P_{8}+P_{10}+P_{11}\right)=\nonumber\\
&\simeq\frac{-\Omega_{0}^{2}\left(\gamma+\mathcal{P}\right)}
{4[(\omega-\omega_{0})^{2}+\Gamma^{2}/4]}\simeq-\Omega^{2}_{0}/\Gamma,\:\:|\omega-\omega_{0}|\ll\Gamma,\label{Ppsi}
\end{align}
yielding $\frac{d}{dt}\left(P_{\Psi_{0}\nrightarrow\Psi_{0}}+P_{\Psi_{0}\rightarrow\Psi_{0}}\right)_{\mathcal{O}(\Omega^{2}_{0}/\Gamma)}=0$, as expected. 
Note that the contributions of diagrams (5) and (9) have been discarded in Eqs.(\ref{Pnpsi}) and (\ref{Ppsi}), respectively, as they are of orders $\gamma_{0}$ 
and $\gamma_{0}\Omega_{0}^{2}/\omega^{2}$ instead. 

\subsection{Semiclassical computation of the coherent scattered power}

\begin{figure}
\begin{center}
\includegraphics[height=60mm,width=70mm,angle=0,clip]{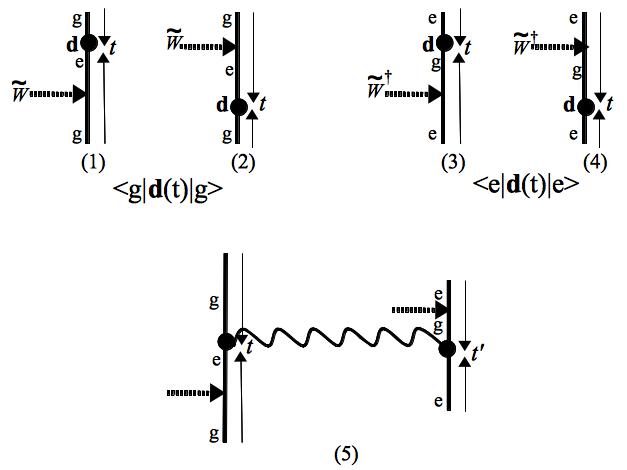}
\caption{Diagrammatic representation of the four processes that contribute, at leading order, to the expectation values $\langle g|\mathbf{d}(t)|g\rangle$ 
[(1) and (2)] and $\langle e|\mathbf{d}(t)|e\rangle$ [(3) and (4)]. Dashed arrows depicts the action of the interaction potential $\tilde{W}$, 
while black solid cyrcles represent the action of the dipole moment operator at the observation time, $t$. In diagram (5) we represent one of the terms of 
$\mathcal{W}_{coh}$ in Eq.(\ref{Wcoh}), proportional to $\gamma\mathcal{P}/\Gamma^{2}$, where several expectation values are combined. The wavy line represents the time propagator of the electric field.}\label{4_diagrams} 
\end{center}
\end{figure}

In semiclassical approaches based on the density functional formalism \cite{Lagendijk_3_level,Scully,Carmichel}  and linear response theory 
\cite{Novotny,ManjavacasPT}, it is customary to refer as coherent scattered power to that in phase with the steady oscillations of the 
expectation value of the atomic dipole moment, $\langle\mathbf{d}(t)\rangle$. In order to interpret the coherent power in terms of our Hamiltonian approach, 
let us consider the probe field as classical. The Hamiltonian of the interaction between the atomic dipole and the probe field comes to depend on time and writes 
$W=\tilde{W}(t)+\tilde{W}^{\dagger}(t)$, with $\tilde{W}=\mathbf{d}\cdot\bm{\epsilon}E_{0}e^{-i\omega t}/2i$, and the probe field photons are to be dropped 
from the steady state of the system in Eq.(\ref{wavefunction}), which becomes now an incoherent superposition of the states
$|\tilde{\Psi}_{0}\rangle_{g}=\sqrt{\gamma/\Gamma}|g\rangle$ and $|\tilde{\Psi}_{0}\rangle_{e}=\sqrt{\mathcal{P}/\Gamma}|e\rangle$. 
Next, let us use a complex-valued representation for the expectation values such that the physical values correspond to 
their real parts. Applying standard time-dependent perturbation theory, the complex-valued averaged expectation value of the total dipole moment reads, 
in the steady state, $\Gamma t\gg1$, 
\begin{align}
\langle\mathbf{d}(t)\rangle&
=\frac{\gamma}{\Gamma}\langle g|\mathbf{d}(t)|g\rangle+\frac{\mathcal{P}}{\Gamma}\langle e|\mathbf{d}(t)|e\rangle\nonumber\\
&=\frac{\mathcal{P}-\gamma}{\Gamma}\frac{ie^{-i\omega t}\Omega_{0}
\boldsymbol{\mu}}{\omega-\omega_{0}+i\Gamma/2}\equiv\langle\mathbf{d}(\omega)\rangle e^{-i\omega t},\label{effectdipole}
\end{align}
where the two terms on the right hand side of the first equality are represented by the diagrams (1) and (4) of  Fig.\ref{4_diagrams}, and off-resonant components have been discarded. 
From Eq.(\ref{effectdipole}) the effective polarizability can be readily identified with \cite{Lagendijk_3_level}
\begin{equation}
\alpha(\omega)=\frac{\mathcal{P}-\gamma}{\Gamma}\frac{\boldsymbol{\mu}\boldsymbol{\mu}}{\hbar(\omega-\omega_{0}+i\Gamma/2)},
\end{equation}
such that $\langle\mathbf{d}(\omega)\rangle=\alpha(\omega)\cdot\bm{\epsilon}E_{0}$. Note that the effective polarizability vanishes for equal population rates, $\mathcal{P}=\gamma$, and so does the expectation value of the dipole moment. 

Further, applying semiclassical linear response theory, the complex-valued coherent field created at position $\mathbf{r}$ and time $t$ by the atomic dipole reads 
\begin{equation}
\mathbf{E}(\mathbf{r},\omega)=-k^{2}\epsilon_{0}^{-1}\mathbb{G}(\mathbf{r}-\mathbf{r}_{A};\omega)\langle\mathbf{d}(\omega)\rangle,\label{Ecoh}
\end{equation}
where $\mathbb{G}(\mathbf{r},\mathbf{r}_{A},\omega)$ is given in Eq.(\ref{Green}), which is the retarded time-Fourier transform of the vacuum commutator of the 
electric field \cite{WyleySipe}, $\mathbb{G}(\mathbf{r}-\mathbf{r}_{A};t-t')\propto i\langle0|[\mathbf{E}(\mathbf{r},t),\mathbf{E}(\mathbf{r}_{A},t')]|0\rangle$, $t\geq t'$.
Correspondingly, the power emitted by the coherent dipole is defined as the time-average rate of the interaction of the induced dipole with its own 
electric field \cite{Novotny},
\begin{align}
\mathcal{W}_{coh}&=\frac{-\omega}{2}\textrm{Im}\:\{\langle\mathbf{d}(\omega)\rangle\cdot\mathbf{E}^{*}(\mathbf{r}_{A},\omega)\}\nonumber\\
&=-\frac{\omega^{3}}{c^{2}\epsilon_{0}}\langle\mathbf{d}(\omega)\rangle\cdot\textrm{Im}\:\mathbb{G}(\mathbf{R};\omega)\cdot\langle\mathbf{d}^{*}(\omega)\rangle\label{Wcoh}\\
&=\frac{-(\mathcal{P}-\gamma)^{2}}{\Gamma^{2}}\frac{\omega^{3}\Omega_{0}^{2}
\boldsymbol{\mu}\cdot\textrm{Im}\:\mathbb{G}(\mathbf{R};\omega)\cdot\boldsymbol{\mu}}{2c^{2}\epsilon_{0}[(\omega-\omega_{0})^{2}+\Gamma^{2}/4]},\quad 
R\rightarrow0,\nonumber
\end{align}
which, oddly enough, vanishes for $\mathcal{P}=\gamma$. Our fully Hamiltonian and quantum computation of Eqs.(\ref{GTgg}), (\ref{GTee}) and (\ref{powerscat}) yields instead 
\begin{equation}
\mathcal{W}_{sc}=-\frac{\omega^{3}\Omega_{0}^{2}
\boldsymbol{\mu}\cdot\textrm{Im}\:\mathbb{G}(\mathbf{R};\omega)\cdot\boldsymbol{\mu}}{2c^{2}\epsilon_{0}[(\omega-\omega_{0})^{2}+\Gamma^{2}/4]},\quad 
R\rightarrow0,
\end{equation}
which differs from the semiclassical calculation of $\mathcal{W}_{coh}$ in the factor $(\mathcal{P}-\gamma)^{2}/\Gamma^{2}$. More importantly, we note that while 
$\mathcal{W}_{sc}$ corresponds to the time-derivative of the quantum expectation value of the electromagnetic energy according to Eq.(\ref{powerscat}),  
$\mathcal{W}_{coh}$ in Eq.(\ref{Wcoh}) does not correspond to the quantum expectation value of any observable, but to the product of several expectation 
values inspired by classical formulas \cite{Novotny}. The fact that no term within $\mathcal{W}_{sc}$ is quadratic in the population rates 
[see Eqs.(\ref{GTgg}) and (\ref{GTee}) in the Appendix] suggests that the 
semiclassical calculation is not a good approximation. Hence, the term in $\mathcal{W}_{coh}$ proportional to $\gamma\mathcal{P}/\Gamma^{2}$ is the result of the 
coupling between $_{g}\langle\tilde{\Psi}_{0}|\mathbf{d}(t)|\tilde{\Psi}_{0}\rangle_{g}$ and $_{e}\langle\tilde{\Psi}_{0}|\mathbf{d}(t)|\tilde{\Psi}_{0}\rangle_{e}$, which are mutually incoherent indeed --see Fig.\ref{4_diagrams}.

\section{Conclusions}\label{lasec6}

Based on QED Hamiltonian perturbation theory, we have developed a wavefunction approach to characterize the 
optical response of a three-level atom subjected to an incoherent pump and illuminated by 
a continuous, weak, and quasi-resonant probe field. In the first place, we integrate adiabatically the dynamics of the upper atomic state,  
and incorporate the incoherent dynamics associated to the pump and the spontaneous emission in exponentially attenuating factors which accompany the hermitian time propagator.  
We have verified that our approach is compatible with unitarity. 
 
We have identified  all the radiative processes which contribute to scattering, absorption and spontaneous emission, and have depicted them diagrammatically. 
We have found that, generically, the pump enhances the linewidth and attenuate the amplitude of the spectra. Besides, extinction differs from that of a free atom in its ground state in the diminishing of the nonradiative losses and in the compensation of the losses by stimulated emission [Eq.(\ref{sigmaext})]. Both effects are proportional to the  steady 
population of the excited state, being the associated energy supplied by the pump. Finally, the incoherent power of Eq.(\ref{powerincoh}), associated to the 
spontaneous decay from the excited to the ground state, is supplied by the pump too. 

Extinction becomes negative for sufficiently strong pumping rate, vanishing for 
$\mathcal{P}=[\gamma_{\omega}+\sqrt{4\gamma_{nr}^{2}+8\gamma_{nr}\gamma_{\omega}+5\gamma_{\omega}^{2}}]/2$. At this point gains and losses compensate,  
satisfying one of the necessary conditions for $\mathcal{PT}$-symmetry in an optical system.

In passing, we have shown that a semiclassical calculation  fails in providing a good estimate of the scattered power.

Our development paves the way for its extension to  many-atom systems with gains and losses. To 
this respect, in a prospective work we plan to characterize the optical response of a pair of identical atoms, with one of them continuously pumped, in order to study its $\mathcal{PT}$-symmetry  properties.

\acknowledgments
We gratefully acknowledge helpful discussions with Alejandro Manjavacas and Julio S\'anchez-C\'anovas. 
This work has been sponsored by Junta de Castilla y Le\'on with Grants VA137G18 and BU229P18.

\appendix

\section{Diagram reading and quantum expressions}

In this Appendix we compile the complete expressions of the quantum processes represented diagrammatically in the main text, in terms of quantum states and operators.  

\begin{widetext}
\subsection{Incoherent transition rates}\label{app1}
The expression corresponding to the diagrams of Fig.\ref{fig2p}(a) for the incoherent pump transition rate of Eq.(\ref{eqP}) reads
\begin{align}
\mathcal{P}&=\frac{\mathop{d}}{\mathop{dt}}\sum_{\mathbf{k}',\bm{\epsilon}'}\Bigl\{..\Bigr\}^{\dagger}\cdot\Bigl\{\hbar^{-2}\int_{0}^{t}d\tau\int_{0}^{\tau}d\tau'
\mathbb{U}_{0}(t-\tau)|1_{\mathbf{k}',\bm{\epsilon}'},(N-1)_{\mathbf{k}_{p},\bm{\epsilon}_{p}};e\rangle\nonumber\\
&\times\langle1_{\mathbf{k}',\bm{\epsilon}'},(N-1)_{\mathbf{k}_{p},\bm{\epsilon}_{p}};e|\mathbf{d}\cdot\mathbf{E}_{\mathbf{k}',\bm{\epsilon}'}^{(-)}
(\mathbf{r}_{A})|(N-1)_{\mathbf{k}_{p},\bm{\epsilon}_{p}};u\rangle
\langle(N-1)_{\mathbf{k}_{p},\bm{\epsilon}_{p}};u|\mathbb{U}_{0}(\tau-\tau')e^{-\gamma_{u}(\tau-\tau')/2}|(N-1)_{\mathbf{k}_{p},\bm{\epsilon}_{p}};u\rangle
\nonumber\\
&\times\langle(N-1)_{\mathbf{k}_{p},\bm{\epsilon}_{p}};u|\mathbf{d}\cdot\mathbf{E}(\mathbf{r}_{A})|N_{\mathbf{k}_{p},\bm{\epsilon}_{p}};g\rangle
\langle N_{\mathbf{k}_{p},\bm{\epsilon}_{p}};g|\mathbb{U}_{0}(\tau')|N_{\mathbf{k}_{p},\bm{\epsilon}_{p}};g\rangle\Bigr\}\nonumber\\
&=\Omega_{p}^{2}\textrm{Re}\:\frac{d}{dt}\int_{0}^{\infty}\frac{-dk'c\:k^{'2}}{\epsilon_{0}\hbar\pi}\hat{\boldsymbol{\mu}}\cdot\textrm{Im}\:\mathbb{G}(\mathbf{R};k')
\cdot\hat{\boldsymbol{\mu}}\Bigl|\int_{0}^{t}d\tau
\:e^{-i(t-\tau)(\omega'+\omega_{0})}\int_{0}^{\tau}d\tau'\:e^{-i(\tau-\tau')\omega_{u}}\:e^{-(\tau-\tau')\gamma_{u}/2}\:e^{-i\tau'\omega_{p}}\Bigr|^{2}\nonumber\\
&\simeq\frac{\Omega_{p}^{2}c}{2\pi\epsilon_{0}\hbar}\textrm{Re}\int_{0}^{\infty}dk'\:i
\frac{k^{'2}\hat{\boldsymbol{\mu}}\cdot\textrm{Im}\:\mathbb{G}(\mathbf{R};k')\cdot\hat{\boldsymbol{\mu}}\:e^{it(\omega'-\omega_{u}+\omega_{0})}}
{(\omega'-\omega_{u}+\omega_{0})[(\omega'-\omega_{u}+\omega_{0})^{2}+\gamma_{u}^{2}/4]},
\quad\gamma_{u}t\gg1,\quad R\rightarrow0^{+},\label{Peq}
\end{align}
where $\Bigl\{..\Bigr\}^{\dagger}$ is the conjugate transpose of the state whose expression appears within curly brackets on its right hand side after the 
dot product symbol. The exponential factor $e^{-\gamma_{u}(\tau-\tau')/2}$ stems from the addition of all the one-photon emission-reabsorption intermediate processes in 
Fig.\ref{fig2p}(a), and the rapid decay from $u$ to $e$ is accounted for by the condition $\gamma_{u}t\gg1$. As for the diagrams of 
Fig.\ref{fig2p}(b) corresponding to the spontaneous emission rate from state $e$, they read
\begin{align}
\gamma_{0}&=\frac{\mathop{d}}{\mathop{dt}}\sum_{\mathbf{k}',\bm{\epsilon}'}\Bigl\{..\Bigr\}^{\dagger}\cdot\Bigl\{\hbar^{-1}\int_{0}^{t}d\tau
\mathbb{U}_{0}(t-\tau)|1_{\mathbf{k}',\bm{\epsilon}'},N_{\mathbf{k}_{p},\bm{\epsilon}_{p}};g\rangle
\langle1_{\mathbf{k}',\bm{\epsilon}'},N_{\mathbf{k}_{p},\bm{\epsilon}_{p}};g|\mathbf{d}\cdot\mathbf{E}_{\mathbf{k}',\bm{\epsilon}'}^{(-)}
(\mathbf{r}_{A})|N_{\mathbf{k}_{p},\bm{\epsilon}_{p}};e\rangle\nonumber\\
&\times\langle N_{\mathbf{k}_{p},\bm{\epsilon}_{p}};e|\mathbb{U}_{0}(\tau)e^{-\gamma\tau/2}|N_{\mathbf{k}_{p},\bm{\epsilon}_{p}};e\rangle
=\frac{d}{dt}\int_{0}^{\infty}\frac{-dk'c\:k^{'2}}{\epsilon_{0}\hbar\pi}\boldsymbol{\mu}\cdot\textrm{Im}\:\mathbb{G}(\mathbf{R};k')
\cdot\boldsymbol{\mu}\Bigl|\int_{0}^{t}d\tau
\:e^{-i(t-\tau)\omega'}\:e^{-i\tau\omega_{0}}\:e^{-\tau\gamma/2}\Bigr|^{2}\nonumber\\
&\simeq\:\int_{0}^{\infty}\frac{dk'c}{\epsilon_{0}\hbar\pi}\frac{k^{'2}\boldsymbol{\mu}\cdot\textrm{Im}\:\mathbb{G}(\mathbf{R};k')\cdot\boldsymbol{\mu}
\left[e^{it(\omega'-\omega_{0})}[\gamma/2-i(\omega'-\omega_{0})]+e^{-it(\omega'-\omega_{0})}
[\gamma/2+i(\omega'-\omega_{0})]-\gamma\right]}
{(\omega'-\omega_{0})^{2}+\gamma^{2}/4},\:\gamma t\ll1,\: R\rightarrow0^{+},\label{goeq}
\end{align}
where the exponential factor $e^{-\gamma\tau/2}$ stems from the addition of all the explicit one-photon emission-reabsorption intermediate processes as well as 
the implicit non-radiative decay channels in  Fig.\ref{fig2p}(b). Slow decay is implicit in the condition $\gamma t\ll1$.

\subsection{Scattered, absorbed, and incoherent emission powers}\label{app2}
The expressions of the scattered power corresponding to the diagrams of Fig.\ref{fig2}(1) and (2) according to Eq.(\ref{powerscat}) read, respectively,
\begin{align}
\mathcal{W}^{(1)}&=\frac{\mathop{d}}{\mathop{dt}}\langle\Psi_{1}(t)|\Psi_{n}^{f}\rangle\langle\Psi_{1}^{f}|H_{EM}|\Psi_{1}^{f}\rangle
\langle\Psi_{1}^{f}|\Psi_{1}(t)\rangle=\frac{\gamma}{\Gamma}\frac{\mathop{d}}{\mathop{dt}}
\sum_{\mathbf{k}',\bm{\epsilon}'}\hbar\omega'\Bigl\{..\Bigr\}^{\dagger}\cdot\Bigl\{\hbar^{-2}\int_{0}^{t}d\tau\int_{0}^{\tau}d\tau'
a_{\mathbf{k}',\bm{\epsilon}'}
\mathbb{U}_{0}(t-\tau)|1_{\mathbf{k}',\bm{\epsilon}'},(N-1)_{\mathbf{k},\bm{\epsilon}};g\rangle\nonumber\\
&\times\langle1_{\mathbf{k}',\bm{\epsilon}'},(N-1)_{\mathbf{k},\bm{\epsilon}};e|\mathbf{d}\cdot\mathbf{E}_{\mathbf{k}',\bm{\epsilon}'}^{(-)}
(\mathbf{r}_{A})|(N-1)_{\mathbf{k},\bm{\epsilon}};e\rangle
\langle(N-1)_{\mathbf{k},\bm{\epsilon}};e|\mathbb{U}_{0}(\tau-\tau')e^{-\Gamma(\tau-\tau')/2}|(N-1)_{\mathbf{k},\bm{\epsilon}};e\rangle
\nonumber\\
&\times\langle(N-1)_{\mathbf{k},\bm{\epsilon}};e|\mathbf{d}\cdot\mathbf{E}(\mathbf{r}_{A})|N_{\mathbf{k},\bm{\epsilon}};g\rangle
\langle N_{\mathbf{k},\bm{\epsilon}};g|\mathbb{U}_{0}(\tau')|N_{\mathbf{k},\bm{\epsilon}};g\rangle\Bigr\}\nonumber\\
&=\frac{\Omega_{0}^{2}\gamma}{\Gamma}\textrm{Re}\:\frac{d}{dt}\int_{0}^{\infty}\frac{-dk'c^{2}k^{'3}}{\epsilon_{0}\pi}
\hat{\boldsymbol{\mu}}\cdot\textrm{Im}\:\mathbb{G}(\mathbf{R};k')
\cdot\hat{\boldsymbol{\mu}}\Bigl|\int_{0}^{t}d\tau
\:e^{-i(t-\tau)\omega'}\int_{0}^{\tau}d\tau'\:e^{-i(\tau-\tau')\omega_{0}}\:e^{-(\tau-\tau')\Gamma/2}\:e^{-i\tau'\omega}\Bigr|^{2}\nonumber\\
&\simeq\frac{-\Omega_{0}^{2}c^{2}\gamma}{2\pi\epsilon_{0}\Gamma}\textrm{Im}\:\int_{0}^{\infty}dk'
\frac{k^{'3}\boldsymbol{\mu}\cdot\textrm{Im}\:\mathbb{G}(\mathbf{R};k')\cdot\boldsymbol{\mu}\:e^{it(\omega'-\omega)}}
{(\omega'-\omega)(\omega'-\omega_{0}-i\Gamma/2)(\omega-\omega_{0}+i\Gamma/2)}=\frac{-\Omega_{0}^{2}c^{2}\gamma}{4\epsilon_{0}\Gamma}\frac{2\omega^{3}\boldsymbol{\mu}\cdot\textrm{Im}
\mathbb{G}(\mathbf{R};\omega)\cdot\boldsymbol{\mu}}{(\omega-\omega_{0})^{2}+\Gamma^{2}/4},
\:\Gamma t\gg1,\:R\rightarrow0^{+},\label{GTgg}
\end{align}
\begin{align}
\mathcal{W}^{(2)}&=\frac{\mathop{d}}{\mathop{dt}}\langle\Psi_{2}(t)|\Psi_{2}^{f}\rangle\langle\Psi_{2}^{f}|H_{EM}|\Psi_{2}^{f}\rangle
\langle\Psi_{2}^{f}|\Psi_{2}(t)\rangle=\frac{\mathcal{P}}{\Gamma}
\frac{\mathop{d}}{\mathop{dt}}\sum_{\mathbf{k}',\bm{\epsilon}'}\hbar\omega'\Bigl\{..\Bigr\}^{\dagger}\cdot\Bigl\{\hbar^{-2}\int_{0}^{t}d\tau\int_{0}^{\tau}d\tau'
a_{\mathbf{k}',\bm{\epsilon}'}\mathbb{U}_{0}(t-\tau)
|1_{\mathbf{k}',\bm{\epsilon}'},(N-1)_{\mathbf{k},\bm{\epsilon}};e\rangle\nonumber\\
&\times\langle1_{\mathbf{k}',\bm{\epsilon}'},(N-1)_{\mathbf{k},\bm{\epsilon}};e|\mathbf{d}\cdot\mathbf{E}(\mathbf{r}_{A})|1_{\mathbf{k}',\bm{\epsilon}'},N_{\mathbf{k},\bm{\epsilon}};g\rangle
\langle1_{\mathbf{k}',\bm{\epsilon}'},N_{\mathbf{k},\bm{\epsilon}};g|\mathbb{U}_{0}(\tau-\tau')
|1_{\mathbf{k}',\bm{\epsilon}'},N_{\mathbf{k},\bm{\epsilon}};g\rangle\nonumber\\
&\times\langle\mathbf{k}',N_{\mathbf{k},\bm{\epsilon}};g|\mathbf{d}\cdot\mathbf{E}_{\mathbf{k}',\bm{\epsilon}'}^{(-)}
(\mathbf{r}_{A})|N_{\mathbf{k},\bm{\epsilon}};e\rangle\langle N_{\mathbf{k},\bm{\epsilon}};e|\mathbb{U}_{0}(\tau')|N_{\mathbf{k},\bm{\epsilon}};e\rangle\Bigr\}
\nonumber\\
&=\frac{\Omega_{0}^{2}\mathcal{P}}{4\Gamma}
\frac{d}{dt}\int_{0}^{\infty}\frac{-dk'c^{2}k^{'3}}{\epsilon_{0}\pi}\boldsymbol{\mu}\cdot\textrm{Im}\:\mathbb{G}(\mathbf{R};k')\cdot\boldsymbol{\mu}\Bigl|\int_{0}^{t}d\tau
\:e^{-i(t-\tau)(\omega'+\omega_{0})}\int_{0}^{\tau}d\tau'\:e^{-i(\tau-\tau')(\omega'+\omega)}\:e^{-(\tau-\tau')\Gamma/2}\:e^{-i\tau'(\omega_{0}+\omega)}\Bigr|^{2}\nonumber\\
&\simeq\frac{-\Omega_{0}^{2}c^{2}\mathcal{P}}{2\pi\epsilon_{0}\Gamma}\textrm{Im}\int_{0}^{\infty}dk'
\frac{k^{'3}\boldsymbol{\mu}\cdot\textrm{Im}\:\mathbb{G}(\mathbf{R};k')\cdot\boldsymbol{\mu}\:e^{it(\omega'-\omega)}}
{(\omega'-\omega)(\omega'-\omega_{0}-i\Gamma/2)(\omega-\omega_{0}+i\Gamma/2)}=\frac{-\Omega_{0}^{2}\mathcal{P}}{4c^{2}\epsilon_{0}\Gamma}\frac{2\omega^{3}\boldsymbol{\mu}\cdot\textrm{Im}
\mathbb{G}(\mathbf{R};\omega)\cdot\boldsymbol{\mu}}{(\omega-\omega_{0})^{2}+\Gamma^{2}/4},
\:\Gamma t\gg1,\:R\rightarrow0^{+}.\label{GTee}
\end{align}
The contributions of diagrams (3) and (4) of Fig.\ref{fig2} to the power absorbed by the atom according to Eq.(\ref{powerabs}) are, respectively,  
\begin{align}
\mathcal{W}^{(3)}&=\Gamma[\langle\Psi_{3}(0)|H_{EM}|\Psi_{3}(0)\rangle
-\langle\Psi_{3}(t)|\Psi_{3}^{f}\rangle\langle\Psi_{3}^{f}|H_{EM}|\Psi_{3}^{f}\rangle\langle\Psi_{3}^{f}|\Psi_{3}(t)\rangle]=
\Gamma\frac{\gamma}{\Gamma}\sum_{\mathbf{k}',\bm{\epsilon}'}\langle N_{\mathbf{k},\bm{\epsilon}};g|\hbar\omega'a^{\dagger}_{\mathbf{k}',\bm{\epsilon}'}
a_{\mathbf{k}',\bm{\epsilon}'}|N_{\mathbf{k},\bm{\epsilon}};g\rangle\nonumber\\
&-\Gamma\frac{\gamma}{\Gamma}\sum_{\mathbf{k}',\bm{\epsilon}'}\hbar\omega'\Bigl\{..\Bigr\}^{\dagger}\cdot
\Bigl\{\hbar^{-1}\int_{0}^{t}d\tau
a_{\mathbf{k}',\bm{\epsilon}'}\mathbb{U}_{0}(t-\tau)e^{-\Gamma(t-\tau)/2}|(N-1)_{\mathbf{k},\bm{\epsilon}};e\rangle\langle (N-1)_{\mathbf{k},\bm{\epsilon}};e|
\mathbf{d}\cdot\mathbf{E}(\mathbf{r}_{A})|N_{\mathbf{k},\bm{\epsilon}};g\rangle\nonumber\\
&\times\langle N_{\mathbf{k},\bm{\epsilon}};g|\mathbb{U}_{0}(\tau)|N_{\mathbf{k},\bm{\epsilon}};g\rangle\Bigr\}=(\hbar\omega\gamma\Omega_{0}^{2}/4)
\Bigl |\int_{0}^{t}d\tau
\:e^{-i(t-\tau)\omega_{0}}\:\:e^{-(t-\tau)\Gamma/2}\:e^{-i\tau\omega}\Bigr|^{2},\label{GT1}
\end{align}
\begin{align}
\mathcal{W}^{(4)}&=\Gamma[\langle\Psi_{4}(0)|H_{EM}|\Psi_{4}(0)\rangle
-\langle\Psi_{4}(t)|\Psi_{4}^{f}\rangle\langle\Psi_{4}^{f}|H_{EM}|\Psi_{4}^{f}\rangle\langle\Psi_{4}^{f}|\Psi_{4}(t)\rangle]=
\Gamma\frac{\mathcal{P}}{\Gamma}\sum_{\mathbf{k}',\bm{\epsilon}'}\langle N_{\mathbf{k},\bm{\epsilon}};e|\hbar\omega'a^{\dagger}_{\mathbf{k}',\bm{\epsilon}'}
a_{\mathbf{k}',\bm{\epsilon}'}|N_{\mathbf{k},\bm{\epsilon}};e\rangle\nonumber\\
&-\Gamma(\mathcal{P}/\Gamma)\sum_{\mathbf{k}',\bm{\epsilon}'}\hbar\omega'\Bigl\{..\Bigr\}^{\dagger}\cdot
\Bigl\{\hbar^{-1}\int_{0}^{t}d\tau
a_{\mathbf{k}',\bm{\epsilon}'}\mathbb{U}_{0}(t-\tau)e^{-\Gamma(t-\tau)/2}|(N+1)_{\mathbf{k},\bm{\epsilon}};g\rangle\langle (N+1)_{\mathbf{k},\bm{\epsilon}};g|
\mathbf{d}\cdot\mathbf{E}(\mathbf{r}_{A})|N_{\mathbf{k},\bm{\epsilon}};e\rangle\nonumber\\
&\times\langle N_{\mathbf{k},\bm{\epsilon}};e|\mathbb{U}_{0}(\tau)|N_{\mathbf{k},\bm{\epsilon}};g\rangle\Bigr\}
=-(\hbar\omega\mathcal{P}\Omega_{0}^{2}/4)
\Bigl |\int_{0}^{t}d\tau
\:e^{-2i(t-\tau)\omega}\:\:e^{-(t-\tau)\Gamma/2}\:e^{-i\tau(\omega+\omega_{0})}\Bigr|^{2}.\label{GT2}
\end{align}
Finally, the contributions of diagrams (5), (6) and (7) of Fig.\ref{fig2} to the incoherent power emitted spontaneously by the atom according to Eq.(\ref{powerincoh}) are, respectively, 
\begin{align}
\mathcal{W}^{(5)}&=\Gamma[\langle\Psi_{5}(t)|\Psi_{5}^{f}\rangle\langle\Psi_{5}^{f}|H_{EM}|\Psi_{5}^{f}\rangle\langle\Psi_{5}^{f}|\Psi_{5}(t)\rangle
-\langle\Psi_{5}(0)|H_{EM}|\Psi_{5}(0)\rangle]=\Gamma\frac{\mathcal{P}}{\Gamma}
\sum_{\mathbf{k}',\bm{\epsilon}'}\hbar\omega'\Bigl\{..\Bigr\}^{\dagger}\nonumber\\
&\cdot\Bigl\{\hbar^{-1}\int_{0}^{t}d\tau
a_{\mathbf{k}',\bm{\epsilon}'}|\mathbb{U}_{0}(t-\tau)e^{-\Gamma(t-\tau)/2}|1_{\mathbf{k}',\bm{\epsilon}'},N_{\mathbf{k},\bm{\epsilon}};g\rangle\langle1_{\mathbf{k}',\bm{\epsilon}'},N_{\mathbf{k},\bm{\epsilon}};g|\mathbf{d}\cdot\mathbf{E}_{\mathbf{k}',\bm{\epsilon}'}^{(-)}
(\mathbf{r}_{A})|N_{\mathbf{k},\bm{\epsilon}};e\rangle
\langle N_{\mathbf{k}_{p},\bm{\epsilon}_{p}};e|\mathbb{U}_{0}(\tau)|N_{\mathbf{k}_{p},\bm{\epsilon}_{p}};e\rangle
\nonumber\\
&=\mathcal{P}\int_{0}^{\infty}\frac{-dk'c^{2}k^{'3}}{\epsilon_{0}\pi}\boldsymbol{\mu}\cdot\textrm{Im}\:\mathbb{G}(\mathbf{R};k')
\cdot\boldsymbol{\mu}\Bigl|\int_{0}^{t}d\tau
\:e^{-i(t-\tau)\omega'}\:e^{-(t-\tau)\Gamma/2}\:e^{-i\tau\omega_{0}}\Bigr|^{2}\nonumber\\
&\simeq-\mathcal{P}\int_{0}^{\infty}\frac{dk'c^{2}}{\epsilon_{0}\pi}\frac{k^{'3}\boldsymbol{\mu}\cdot\textrm{Im}\:\mathbb{G}(\mathbf{R};k')\cdot\boldsymbol{\mu}}
{(\omega'-\omega_{0})^{2}+\Gamma^{2}/4},\quad\Gamma t\ll1,\quad R\rightarrow0^{+},\label{goeqo}
\end{align}
\begin{align}
\mathcal{W}^{(6)}&+\mathcal{W}^{(7)}=\Gamma\sum_{n=6}^{7}\Bigl[\langle\Psi_{n}(t)|\Psi_{n}^{f}\rangle\langle\Psi_{n}^{f}|H_{EM}|\Psi_{n}^{f}\rangle
\langle\Psi_{n}^{f}|\Psi_{n}(t)\rangle-\langle\Psi_{n}(0)|H_{EM}|\Psi_{n}(0)\rangle\Bigr]=
\Gamma\frac{-2\mathcal{P}}{\Gamma}\textrm{Re}\sum_{\mathbf{k}',\bm{\epsilon}'}\hbar\omega'\nonumber\\
&\times\hbar^{-4}\int_{0}^{t}d\xi\langle N_{\mathbf{k},\bm{\epsilon}};e|\mathbb{U}_{0}^{\dagger}(\xi)| N_{\mathbf{k},\bm{\epsilon}};e\rangle
\langle N_{\mathbf{k},\bm{\epsilon}};e|\mathbf{d}\cdot\mathbf{E}_{\mathbf{k}',\bm{\epsilon}'}^{(+)}(\mathbf{r}_{A})
|1_{\mathbf{k}',\bm{\epsilon}'},N_{\mathbf{k},\bm{\epsilon}};g\rangle\langle1_{\mathbf{k}',\bm{\epsilon}'},N_{\mathbf{k},\bm{\epsilon}};g|
a^{\dagger}_{\mathbf{k}',\bm{\epsilon}'}\mathbb{U}_{0}^{\dagger}(t-\xi)e^{-\Gamma(t-\xi)/2}\nonumber\\
&\times\int_{0}^{t}d\tau\int_{0}^{\tau}d\tau'\int_{0}^{\tau'}d\tau''\:a_{\mathbf{k}',\bm{\epsilon}'}
\mathbb{U}_{0}(t-\tau)e^{-\Gamma(t-\tau)/2}|1_{\mathbf{k}',\bm{\epsilon}'},N_{\mathbf{k},\bm{\epsilon}};g\rangle\langle1_{\mathbf{k}',\bm{\epsilon}'},N_{\mathbf{k},\bm{\epsilon}};g|
\mathbf{d}\cdot\mathbf{E}(\mathbf{r}_{A})|1_{\mathbf{k}',\bm{\epsilon}'},(N-1)_{\mathbf{k},\bm{\epsilon}};e\rangle\nonumber\\
&\times\langle1_{\mathbf{k}',\bm{\epsilon}'},(N-1)_{\mathbf{k},\bm{\epsilon}};e|\mathbb{U}_{0}(\tau-\tau')
|1_{\mathbf{k}',\bm{\epsilon}'},(N-1)_{\mathbf{k},\bm{\epsilon}};e\rangle
\langle1_{\mathbf{k}',\bm{\epsilon}'},(N-1)_{\mathbf{k},\bm{\epsilon}};e|\mathbf{d}\cdot\mathbf{E}(\mathbf{r}_{A})|
1_{\mathbf{k}',\bm{\epsilon}'},N_{\mathbf{k},\bm{\epsilon}};g\rangle\nonumber\\
&\times\langle1_{\mathbf{k}',\bm{\epsilon}'},N_{\mathbf{k},\bm{\epsilon}};g|\mathbb{U}_{0}(\tau'-\tau'')e^{-\Gamma(\tau'-\tau'')/2}
|1_{\mathbf{k}',\bm{\epsilon}'},N_{\mathbf{k},\bm{\epsilon}};g\rangle
\langle1_{\mathbf{k}',\bm{\epsilon}'},N_{\mathbf{k},\bm{\epsilon}};g|\mathbf{d}\cdot\mathbf{E}_{\mathbf{k}',\bm{\epsilon}'}^{(-)}
(\mathbf{r}_{A})|N_{\mathbf{k},\bm{\epsilon}};e\rangle\langle N_{\mathbf{k},\bm{\epsilon}};e|\mathbb{U}_{0}(\tau'')|N_{\mathbf{k},\bm{\epsilon}};e\rangle\nonumber\\
&=\Omega_{0}^{2}\mathcal{P}\textrm{Re}\int_{0}^{\infty}\frac{-dk'c^{2}k^{'3}}{2\pi\epsilon_{0}}\boldsymbol{\mu}\cdot\textrm{Im}\:\mathbb{G}(\mathbf{R};k')
\cdot\boldsymbol{\mu}\int_{0}^{t}d\tau\:
e^{-i(t-\tau)(\omega'+\omega)}e^{-\Gamma(t-\tau)/2}\int_{0}^{\tau}d\tau'\:e^{-i(\tau-\tau')(\omega'+\omega_{0})}\:e^{-(\tau-\tau')\Gamma/2}\nonumber\\
&\times\int_{0}^{\tau'}d\tau''\:e^{-i(\tau'-\tau'')(\omega'+\omega)}e^{-(\tau'-\tau'')\Gamma/2}e^{-i\tau''(\omega+\omega_{0})}\int_{0}^{t}d\xi\:
e^{i(t-\xi)(\omega'+\omega)}e^{-\Gamma(t-\xi)/2}e^{i\xi(\omega+\omega_{0})}\nonumber\\
&\simeq\frac{\Omega_{0}^{2}\mathcal{P}c^{2}}{2\pi\epsilon_{0}}\textrm{Re}\int_{0}^{\infty}dk'
k^{'3}\boldsymbol{\mu}\cdot\textrm{Im}\:\mathbb{G}(\mathbf{R};k')\cdot\boldsymbol{\mu}\frac{e^{-it(\omega'-\omega)}-1}
{(\omega'-\omega)(\omega'-\omega_{0}-i\Gamma/2)^{2}(\omega'-\omega_{0}+i\Gamma/2)},
\:\Gamma t\gg1,\:R\rightarrow0^{+}.\label{goeqoo}
\end{align} 
\end{widetext}


\begin{thebibliography}{105}

\bibitem{Sakurai} J.J. Sakurai, \textit{Advanced Quantum Mechanics}, Addison-Wesley Publishing Company, Boston (1967).
\bibitem{Milonni_book}  P.W. Milonni,  \textit{The Quantum Vacuum}, Academic Press, San Diego (1994).
\bibitem{Peskin_book} M.E. Peskin and D.V. Schroeder, \textit{An Introduction to Quantum Field Theory}, CRC Press, Taylor and Francis Group, New York (2018).
\bibitem{vanTiggelen} B. A. van Tiggelen, \emph{Mesoscopic light scattering in atomic physics}, in \emph{Coherent atomic matter waves}, edited by R. Kaiser, C. Westbrook, and F. David (Springer Berlin Heidelberg, 2001), pp. 371-414.
\bibitem{Milonni-Berman} P.R. Berman, R. W. Boyd, and P. W. Milonni, Phys. Rev. A\textbf{74}, 053816 (2006).
\bibitem{Loudon-Berman}  P.W. Milonni, R. Loudon, P. R. Berman, and S. M. Barnett, Phys. Rev. A\textbf{77}, 043835 (2008).
\bibitem{Buhmann-Scheel} S. Y. Buhmann, H. T. Dung, T. Kampf, and D. G. Welsch, Eur.Phys. J. D\textbf{35}, 15 (2005).
\bibitem{Coevorden} P. de Vries, D.V. van Coevorden, and A.Lagendijk, Rev. Mod. Phys. \textbf{70}, 447 (1998).
\bibitem{Donaire} M. Donaire, Phys. Rev. A\textbf{83}, 022502 (2011); Phys. Rev. A\textbf{85}, 052518 (2012).
\bibitem{WyleySipe} J.M. Wylie, J.E. Sipe, Phys. Rev. A\textbf{30}, 1185 (1984).
\bibitem{Novotny} L. Novotny, B. Hecht, \textit{Principles of nano-optics}, Cambridge University Press (2006).
\bibitem{OBrien} C. O'Brien, P.M. Anisimov, Y. Rostovtsev, and Olga Kocharovskaya, \emph{Coherent control of refractive index in far-detuned $\Lambda$ systems}, Phys. Rev. A\textbf{84}, 063835 (2011).
\bibitem{Hang} C. Hang, G. Huang, and V.V. Konotop, \emph{$\mathcal{PT}$ Symmetry with a System of Three-Level Atoms}, Phys. Rev. Lett. \textbf{110}, 083604 (2013).
\bibitem{Scully} M. O. Scully, M. S. Zubairy, \emph{Quantum Optics}, Cambridge University Press (1997).
\bibitem{Heyne-Bullough} F. Hynne and R. K. Bullough, Phil. Trans. R. Soc. A\textbf{312}, 251 (1984); \textbf{321}, 305 (1987); \textbf{330}, 253 (1990).
\bibitem{Bullough} R. K. Bullough, J. Phys. A\textbf{1}, 409 (1968); \textbf{2}, 477 (1969); \textbf{3}, 708 (1970); \textbf{3}, 726 (1970); \textbf{3},
751 (1970).
\bibitem{Agarwal} G. S. Agarwal, Phys. Rev. A\textbf{11}, 253 (1975).
\bibitem{Buhmann-Wess} S. Y. Buhmann, H. Safari, D.-G. Welsch, H. Trung Dung, Open Syst. Inf. Dyn. \textbf{13}, 427 (2006).
\bibitem{vanTiggelen-Lagendij_Phys_Rept} A.Lagendijk, B.van Tiggelen, Phys.Rep. \textbf{270}, 143 (1996).
\bibitem{deVries-Lagendij} P. de Vries, D.V. van Coevorden, A.Lagendijk, Rev. Mod. Phys. \textbf{70}, 447 (1998).
\bibitem{Salam} A. Salam, Intermolecular interactions in a radiation field via the method of induced moments, Phys. Rev. A\textbf{73}, 013406 (2006).

\bibitem{Khandekar} W. Jin, C. Khandekar, A. Pick, A.G. Polimeridis, A.W. Rodriguez, \emph{Amplified and directional spontaneous emission from arbitrary
composite bodies: A self-consistent treatment of Purcell effect below threshold}, Phys. Rev. B: Condens. Matter Mater. Phys. \textbf{93}, 125415 (2016).
\bibitem{ManjavacasPT} A. Manjavacas, \emph{Anisotropic optical response of nanostructures with balanced gain and loss}, ACS Photonics \textbf{3}, 1301 (2016).

\bibitem{TejedorII} Elena del Valle, Fabrice P. Laussy, and Carlos Tejedor, \emph{Luminescence spectra of quantum dots in microcavities. I. Bosons}, Phys. Rev. B\textbf{79}, 235325 (2009); \emph{Luminescence spectra of quantum dots in microcavities. II. Fermions}, Phys. Rev. B\textbf{79}, 235326 (2009).
\bibitem{German} G.J. de Valc\'arcel, E. Rold\'an, F. Prati, Semiclassical theory of amplification and lasing, \textit{Revista Mexicana de F\'isica} E\textbf{52}, 
198 (2006).
\bibitem{Lagendijk_3_level} Tom Savels, Allard P. Mosk, and Ad Lagendijk, \emph{Light scattering from three-level systems: The T matrix of a point dipole with gain},
Phys. Rev. A 71, 043814 (2005).

\bibitem{Carmichel} H. J. Carmichael. \emph{Statistical Methods in Quantum Optics 1}, Springer, Berlin Heidelberg (1999).
\bibitem{Harry_Paul} H. Pau, \emph{Introduction to Quantum Optics, from Light Quanta to Quantum Teleportation}, Cambridge University Press (2004).
\bibitem{Hertel-Schulz}  I. V. Hertel and C. P. Schulz,  \textit{Atoms, Molecules and Optical Physics 2}, Springer, New York (2015).

\bibitem{footnote} In some particular scenarios, however, within the density functional formalism, the photonic dynamics is also addressed. That is the case of, for instance, 
the dressed-atom formalism for an atom interacting with a strong  probe field \cite{Cohen_Reynaud}, or the Jaynes-Cummings model in cavity-QED, where the spectrum of cavity modes is a subject of study \cite{TejedorII}.
\bibitem{Cohen_Reynaud} C Cohen-Tannoudji and S Reynaud,\emph{Dressed-atom description of resonance
fluorescence and absorption spectra of a multi-level atom in an intense laser beam}, J. Phys. B: Atom. Mol. Phys. 10, 345 (1977).
\bibitem{Bender} \emph{$\mathcal{PT}$-symmetric quantum theory}, J. Phys.: Conf. Ser. \textbf{631}, 012002 (2015).


\bibitem{Japs}  Z. Zhang, Y. Zhang, J. Sheng, L. Yang, M.-A. Miri, D. N.
Christodoulides, B. He, Y. Zhang, and M. Xiao, \emph{Observation of parity-time symmetry in optically induced atomic lattices}, Phys. Rev. Lett. \textbf{117}, 123601 (2016); Z. Zhang, D. Ma, J. Sheng, Y. Zhang, Y. Zhang, and M. Xiao, \emph{Non-hermitian optics in atomic systems}, J. Phys. B \textbf{51}, 072001 (2018).
\bibitem{ORN19} S. K. Ozdemir, S. Rotter, F. Nori, and L. Yang, \emph{Parity- time symmetry and exceptional points in photonics,} Nat. Mater. \textbf{18}, 783 (2019).
\bibitem{JKP16} S. Sanders and A. Manjavacas, \emph{Nanoantennas with balanced gain and loss}, Nanophotonics \textbf{9}, 473 (2020); 
M.-A. Miri, M. A. Eftekhar, M. Facao, A. F. Abouraddy, A. Bakry, M. A. N. Razvi, A. Alshahrie, A. Al\'u, and D. N. Christodoulides, \emph{Scattering properties of pt-symmetric objects}, J. Opt. \textbf{18}, 075104 (2016); 
A. Krasnok, D. Baranov, H. Li, M.-A. Miri, F. Monticone, and A. Al\'u, \emph{Anomalies in light scattering}, Adv. Opt. Photon. \textbf{11}, 892 (2019); 
Y. J. Zhang, P. Li, V. Galdi, M. S. Tong, and A. Al\'u, \emph{Manipulating the scattering pattern with non-hermitian particle arrays}, Opt. Express \textbf{28}, 19492 (2020); 
R. Kolkowski and A. F. Koenderink, \emph{Gain-induced scattering anomalies of diffractive metasurfaces}, Nanophotonics , 20200253 (2020).



\end{thebibliography}
\end{document}